\documentclass{article}
\usepackage[english]{babel}
\usepackage[latin1]{inputenc}
\usepackage{graphicx}
\usepackage{amssymb,amsmath}
\usepackage{pstricks}
\usepackage{geometry}
\def\rt{R_{\rm t}}
\def\ggb{g_{\rm gb}}
\def\egb{\epsilon_{\rm gb}}
\title{Stochastic analysis of the time evolution of Laminar-Turbulent bands of plane Couette flow}
\author{Joran Rolland\footnote{LadHyX, UMR 7646 CNRS, Palaiseau 91128 France, \emph{current address:} Insitut f\"ur Atmosph\"are und Umwelt, Goethe Universit\"at, 60438 Frankfurt, Germany {\sc rolland@iau.uni-frankfurt.de}}}
\date{\today}
\begin{document}

\maketitle

\begin{abstract}
This article is concerned with the time evolution of the oblique laminar-turbulent bands of transitional plane Couette flow under the influence of turbulent noise. Our study is focused on the amplitude of modulation of turbulence (the bands).
In order to guide the numerical study of the flow, we first perform an analytical and numerical analysis of a Stochastic Ginzburg--Landau (GL) equation for a complex order parameter. The modulus of this order parameter models the amplitude of modulation of turbulence. Firstly, we compute the autocorrelation function of said modulus once the band is established. Secondly, we perform a calculation of average and fluctuations around the exponential growth of the order parameter. This type of analysis is similar to the Stochastic Structural Stability Theory (S3T). We then perform numerical simulations of the Navier--Stokes equations in order to confront these predictions with the actual behaviour of the bands. Computation of the autocorrelation function of the modulation of turbulence shows quantitative agreement with the model: in the established band regime, the amplitude of modulation follows an Ornstein--Uhlenbeck process. In order to test the S3T predictions, we perform quench experiments, sudden decreases of the Reynolds number from uniform turbulence, in which modulation appears. We compute the average evolution of the amplitude of modulation and the fluctuations around it. We find good agreement between numerics and modeling. The average trajectory grows exponentially, at a rate clearly smaller than that of the formation of laminar holes. Meanwhile, the actual time evolution remains in a flaring envelope, centered on the average, and expanding at the same rate. These results provide further validation of the stochastic modeling for the time evolution of the bands for further studies. Besides, they stress on the difference between the oblique band formation and the formation of laminar holes.
\end{abstract}

\begin{flushleft}
Shear turbulence, 47.27.nb -- Transition to turbulence, 47.27.Cn -- Stochastic analysis methods, 05.10.Gg
\end{flushleft}

\section{Introduction}

\begin{figure}
\centerline{\textbf{(a)}\includegraphics[width=4cm,clip]{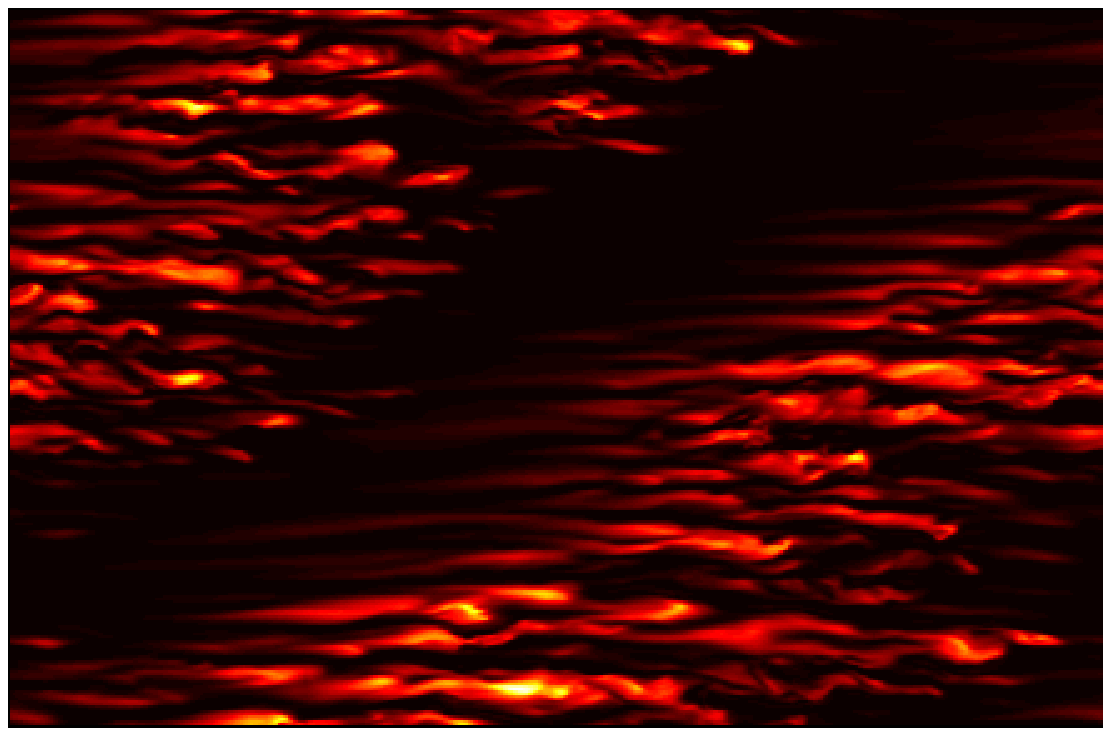}\textbf{(b)}\includegraphics[width=6cm]{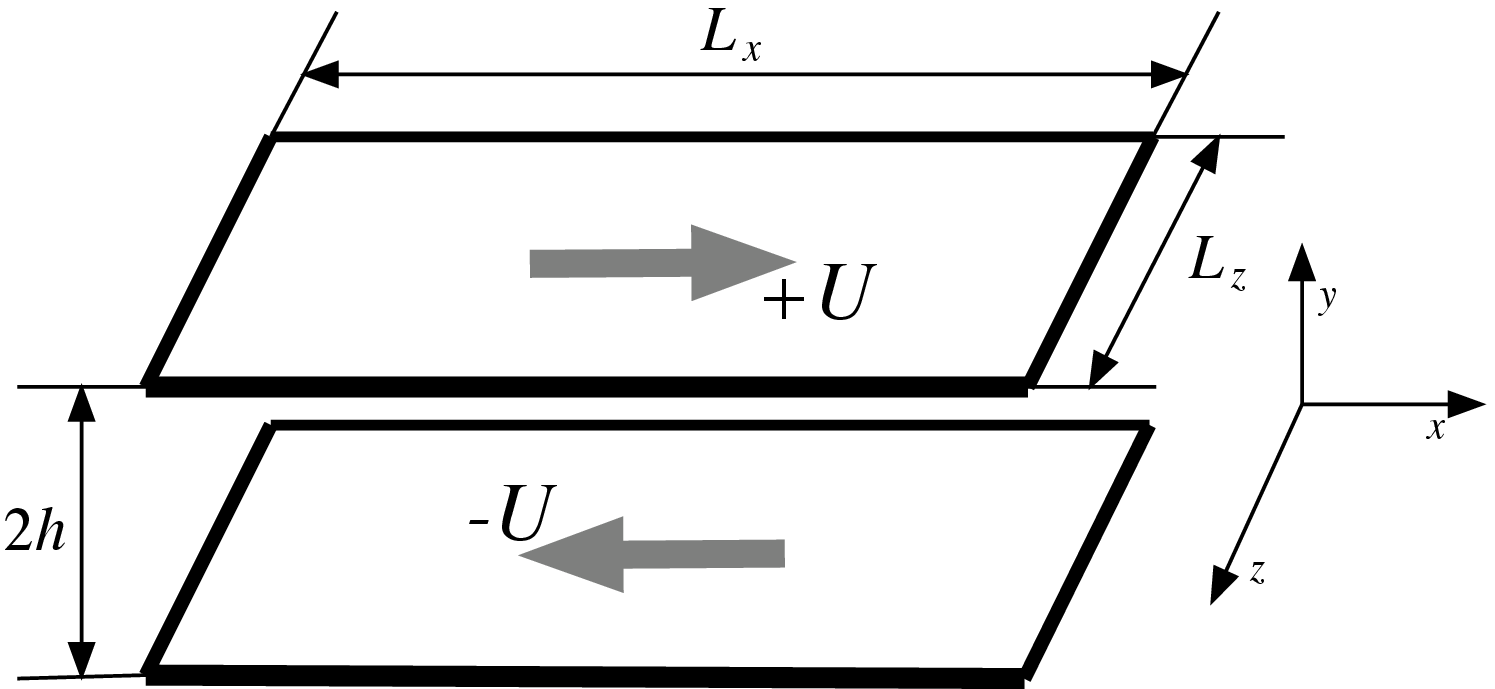}}
\caption{(a) : Colour levels of the norm of the velocity field in the band regime of plane Couette flow (numerical result), in a horizontal plane. (b) : Sketch of the plane Couette flow configuration. The wall normal direction $\vec e_y$ sets the vertical.}
\label{fig1}
\end{figure}

  This article studies the random time evolution of the modulation of turbulence in transitional plane Couette flow (Fig.~\ref{fig1} (a)).
Plane Couette flow is the flow between two parallel plates moving in opposite directions and separated by a constant gap (Fig.~\ref{fig1} (b)). This flow
is linearly stable for all Reynolds numbers $R$, the control parameter. As a consequence, the transition to turbulence is discontinuous :
turbulence requires finite amplitude perturbations to be triggered. Besides, turbulent flow can coexist in space and time with laminar flow,
provided that $R>R_{\rm g}$, the global threshold of transition. There are several manner to define this threshold: $R_{\rm g}$ can be the Reynolds number under which turbulence is not sustained permanently \cite{prigent02,PM}, or more precisely, the Reynolds number under which the mean lifetime of a turbulent germ equals the mean time elapsed before that germ splits \cite{shi}. The most intriguing property of the flow is that up to $R_{\rm t}>R_{\rm g}$, turbulence is sustained, but does not invade the whole domain.
Instead, it takes the form of regular apparently quasi steady oblique laminar-turbulent bands \cite{prigent02,PM,dsc10}. Above $R_{\rm t}$, turbulence occupies the whole domain: again, this Reynolds number has several definitions, all related to the disappearance of the laminar troughs \cite{prigent02,PM,BT11,RM10_1}. Note that the bands can have the two orientations (termed $+$ and $-$) with equal probability. These bands correspond to a sinusoidal modulation of the amplitude of turbulence \cite{BT11,RM10_1}. Said modulation disappears in a very intermittent manner near $R_{\rm t}$ \cite{RM10_1,BT05}, which may in fact be very similar to a critical phenomenon \cite{RM10_1,RNL}. Understanding the type of noise felt by the band is fundamental in explaining to what extent this similarity goes.

There have been many attempts at understanding the hydrodynamic mechanisms sustaining the bands. Advection of small scale structures (size of the gap) by a large scale flow (size of the bands) is central in each of these discussions \cite{PRL,ispspot,BT07}. The topic is still open, since the effect of these small scale structures on the large scale flow has not been precisely explained. Comparatively, the modeling of the laminar-turbulent coexistence is more advanced. Model equations have been proposed that describe the amplitude and localisation of turbulence \cite{bp} or the amplitude of modulation of turbulence $m$ (the bands) \cite{prigent02,RM10_1}. Said amplitude of modulation is usually computed from a Fourier transform of the velocity field \cite{RM10_1,BT11}. In both cases, it is very clear that fluctuating quantities with well defined PDFs and moments are the observables of interest \cite{BT11,RM10_1}.

The time average of the modulation of turbulence (or the most probable value \cite{BT11}) shares many properties of pattern formation or of the average of an order parameter in a critical phenomenon \cite{prigent02,RM10_1}. This led to the proposal of Ginzburg--Landau types of models, for an order parameter $A$, in order to describe the deterministic part of the evolution of the amplitude of modulation of turbulence $m$. The modeling of its fluctuations, and the noise term that should be added to these evolution equations, are not that simple.

Indeed, in the case of the amplitude of turbulence, it is clear that the noise level felt is a growing function of said amplitude, and that both go to zero together. Several types of power law dependencies have been proposed \cite{bp,cras}. This fact is consistent with the assertion that transition to turbulence in shear flows is extremely similar to direct percolation (DP) \cite{shi}. Indeed, in the Langevin description of reaction diffusion models such as DP, the correlations of the noise are proportional to the fields amplitudes \cite{car1}. These Langevin equations are derived from the field theories constructed to study the phase transitions in such models \cite{car2}. Meanwhile, some authors proposed additive gaussian stochastic forcing for the modeling of the small scale rolls and streaks \cite{S3T}. Note, however, that the covariance of their noise is directly dependent on the mean kinetic energy.

In the case of the amplitude of modulation of turbulence, additive white noise has been included in the first model equation for the band on qualitative grounds \cite{prigent02,phD}. The comparison with experimental results is satisfactory. The physical idea motivating this model is that the noise amplitude is still related to the amplitude of turbulence, which can be thought of as a constant when considering the amplitude of modulation of turbulence. This assumes of course that $R$ is not too close to $R_{\rm g}$. This type of modeling has been tested more quantitatively when considering the random walk of the relative position of the band \cite{RM10_1}, or the orientation fluctuations \cite{RM10_2}, in numerical simulations. However, only the later study revealed the the fluctuations of the amplitude of modulation. The range of Reynolds numbers over which the lifetime in one orientation could be compared to the Arhenius type law predicted by the model was not large enough for the noise modeling to be entirely validated.

  The precise description of the type of noise felt by the amplitude of modulation is fundamental to the understanding of the disappearance of the modulation at $R_{\rm t}$. In particular, the assertion that this phenomenon could be described by a critical phenomenon at equilibrium depends on the noise being additive and not multiplicative \cite{RNL}. Therefore, in order to fully justify the use of stochastic amplitude equation models for the laminar turbulent coexistence, each of the prediction of said model must be compared to numerical or experimental data.

  In this article, we will endeavour to test said predictions on the temporal evolution of the modulation of turbulence under its own noise. We will take two points of views: that of the fluctuations of well established bands around their average, and the noisy growth of bands from uniform turbulence. Indeed, if one excludes the case of switching of orientation, the bands only evolve in these two situations. In order to do so, we proceed in the following manner. In section~\ref{sgl}, we remind the stochastic Ginzburg--Landau model, which we analyse analytically and numerically in the two cases of interest. In section~\ref{num}, we present our numerical procedure, and the processing of numerical data. We then analyse the numerical simulations in section~\ref{res} and compare the results to theory. We eventually discuss the results altogether in the conclusion (\S~\ref{conc}).

\section{The Stochastic Ginzburg--Landau model \label{sgl}}

\subsection{Principle of the model\label{ssp}}

\subsubsection{The equation }

 We introduce the stochastic Landau limit of the Ginzburg--Landau--Langevin models used to describe the oblique bands in
Taylor--Couette flow \cite{prigent02,phD} and plane Couette flow \cite{RM10_1,RM10_2}.
This model describes the evolution of two complex fields $A_\pm=A^r_\pm+\imath A^i_\pm$ and reads:
 \begin{equation}
\tau_0 \partial_t A_\pm=\left(\epsilon-g_1|A_\pm|^2-g_2|A_\mp|^2\right)A_\pm+\xi_x^2\partial_x^2A_\pm+\xi_z^2\partial_z^2A_\pm+(\bar{\zeta}_\pm^{\rm r}(x,z,t)+\imath \zeta_\pm^{\rm i}(x,z,t)) \label{GLL}\,,
 \end{equation}
where $\tau_0$ is the time scale, $\xi_{x,z}$ are the correlation lengths, $g_{1,2}$ are positive inverse saturation amplitudes and $\bar{\zeta}_{\pm}^{\rm i,r}$ is a white noise in time and space of variance $\bar{\alpha}^2$ \cite{gar}:
\begin{equation}\langle \bar{\zeta}_i^j(x,z,t)\bar{\zeta}_k^l(x',z',t')\rangle=\bar{\alpha}^2\delta(x-x')\delta(t-t')\delta(z-z')\delta_{ik}\delta_{jl}\,.\end{equation} We use brackets $\langle .\rangle$ for the ensemble averages. The moduli $|A_\pm|$ describe the amplitude of modulation of turbulence at the corresponding orientation $m_\pm$. The phases describe their relative positions in the flow.

This type of equation, with $\bar{\alpha}=0$, arises in weakly non-linear instabilities, particularly in pattern formation, when one performs a projection on the central manifold, that is, an elimination of the fast evolving space and time scales \cite{CH}. This type of equation arises in phase transitions as well, with $\bar{\alpha}=\sqrt{2k_BT}$, $T$ being the temperature and $k_B$ the Boltzmann constant, where it is often termed Model A (gradient system without a conservation law for $A_\pm$) \cite{HA}.

The actual framework in which the band falls is unclear, although it contains elements from these two fields of physics. Indeed, even though it is an hydrodynamical phenomenon, the mechanisms of sustainment appear to be rather different from classical pattern formation instabilities \cite{PRL,ispspot,BT07}. No derivation from first hydrodynamical principles has been proposed. The type of coexistence, quasi steady bands or generally intermittent, depends a lot on the size of the domain \cite{PM}.  The use of the Ginzburg--Landau model to describe the bands is therefore in the same line of thought as the use of the Landau theory to describe phase transitions. The equation is obtained through an experience of thought: a ``coarse graining'': this is an average over the fast spatial scales of the actual hydrodynamics. The equation is then used to describe and understand the main physical properties of the system. Indeed, the formal resemblance between amplitude of modulation $m_\pm$ and $A_\pm$: slow spatial variation, symmetries, noisy signal, phase invariance, no superposition of $+/-$ fields, continuous apparition at $\epsilon=0$, justified the proposition of such a model to match $m_\pm$, with $\epsilon\equiv1-R/R_{\rm t}$ \cite{prigent02,phD}. Many experimental and numerical data were perfectly matched by the model, and gave values for constants $g_1$, $g_2$ and $\xi_z$, ranges for $\tau_0$  and estimates of $\alpha$.

The strength of such stochastic models is that they are easily analysed for large scale systems \cite{bp}. We can make predictions that can then be verified in laboratory or numerical experiments. In the case of plane Couette flow, the timescales of the competitions between domains of $+$ and $-$ orientation (see \cite{DSH}) may very well have time scales deriving from coarsening dynamics (as noted in \cite{EPJB}). Said dynamics are typical of competitions between domains of $+$ and $-$ in models like the GL equation (Eq.~(\ref{GLL})) \cite{bray}. More surprisingly, the very intermittent behaviour at the disappearance of modulation may be very similar to a critical phenomenon \cite{RNL}.

\subsubsection{Evolution equation of the modulus}

Since our DNS are performed in a domain containing a few wavelengths of the band, we move to the Landau version of the Ginzburg--Landau--Langevin model \cite{RM10_1}:
\begin{equation} \tau_0  \partial_t A_\pm
=\underbrace{\left(\epsilon-\xi_x^2\delta k_x^2-\xi_z^2\delta
k_z^2\right)}_{\tilde{\epsilon}} A_\pm
-g_1|A_\pm|^2A_\pm-g_2|A_\mp|^2A_\pm+(\zeta^{\rm r}_\pm(t)+\imath \bar{\zeta}_\pm^{\rm i})\,.
\label{eq10} \end{equation}
The main difference with model~(\ref{GLL}) lies in the fact that $A_\pm$ is now only a function of time. The spatial derivatives are replaced by the distances to the optimal wavenumber of the band: $\delta k_{x,z}=k_{x,z}-k_{x,z}^{\rm c}$, with $k_{x,z}^{\rm c}$ both depending on $R$ \cite{prigent02,RM10_1}. This modification is the same as the one performed in pattern formation, when a modulation of wavelength $\lambda_c=2\pi/k_c$ arises \cite{CH}. There is one subtle difference: this equation is derived from a spatial average of equation~\ref{GLL}, the noise is changed from $\bar{\zeta}(x,t)$ with variance $\bar{\alpha}^2$ to $\zeta(t)$ with variance $\alpha^2=\bar{\alpha}^2/(L_xL_z)$. This type of noise forcing should be expected for spatially averaged quantities. For instance, in the case of the average kinetic energy, this is consistent with the finding that their probability density functions $\rho_E(E)$ have Large Deviations in the limit of large size \cite{EPJB} :
\begin{equation}
-\frac{1}{L_xL_z}\ln(\rho_E)=I(E)\,.\label{LDP}
\end{equation}
The function $I(E)$ is termed a large deviation function \cite{ht}. It appears to be a parabola in a very large range of Reynolds numbers, making $E$ a gaussian random variable.

In a statistically steady state, one finds one solution of equation~(\ref{eq10}) of spatial and ensemble average $\langle A_\pm\rangle =0+O(\alpha)$ if $\tilde{\epsilon} <0$. This solution describes uniform turbulence. One finds two solutions $\langle|A_\pm|\rangle = \sqrt{\frac{\epsilon}{g_1}}+O(\alpha)$, $A_\mp =0+O(\alpha)$ provided $g_2>g_1$ if $\tilde{\epsilon} >0$. These two solutions describe a flow where one orientation of the band is dominant and small traces of the other orientation are found, due to turbulent noise \cite{RM10_1}.

This model can be written using the potential $\bar{V}$:
\begin{equation} \bar{V}=-\frac{\tilde{\epsilon}}{2}\left(|A_+|^2+|A_-|^2
\right)+\frac{g_1}{4}\left(|A_+|^4+|A_+|^4
\right)+\frac{g_2}{2}|A_+|^2|A_-|^2 \label{eq12}\,.\end{equation}
The interest of a potential approach lies in the mean field treatment of the problem: \emph{i.e.} the assimilation of the average to the most likely value when relative fluctuations are small \cite{RM10_1,LL}.

If one is concerned by the dynamics of the modulus $A$ of only one orientation, because the other is negligible or the interactions are weak, the last non-linear coupling term can be neglected. In order to obtain the equation describing the time evolution of the modulus of $A$, one must make a change of variable
$(A_r,A_i)\rightarrow A,\phi$ (dropping the $\pm$). It is non-linear and implies some subtleties. Indeed, a stochastic process of the form:
\begin{equation}\partial_t x=F(x)+G(x)\zeta(t)\,,\end{equation}
with $\zeta$ a white noise, can have several meaning which are explicit only when time is discretised. The most common definition is termed the It\^o rule, where the discretised equation reads:
\begin{equation}x(t+dt)-x(t)=F(x(t))+G(x(t))(\zeta(t+dt)-\zeta(t))\,.\end{equation}
One can demonstrate that performing a change of variables in the same manner as for an ordinary differential equation yields a different physical process (see for instance \cite{gar} for a detailed discussion).

In order to perform the change of variables: a first strategy to change variables can be
deriving the Fokker--Plank equation for the modulus and phase, and then deducing the Langevin equation for the modulus
with a It\^o rule. One can alternatively use the It\^o formula directly in the Langevin equation (see \cite{gar} \S~4.5.5 for a similar derivation).
Both yields an over damped Langevin Equation:
\begin{equation}
 \tau_0 \partial_t A=-\partial_A V+\frac{\alpha^2}{2A}+\zeta_2(t) \,, \,
 V=-\frac{\tilde{\epsilon}}{2}A^2+\frac{g_1}{4}A^4\,,\label{eqchvar}
\end{equation}
with $\zeta_2$ a Gaussian white noise of variance $\alpha^2$. The main consequence of the non-linear change of variable is the introduction of the drift $\frac{\alpha^2}{2A}$, which prevents the modulus from reaching $0$, a property of the modulus of a complex random variable. This is the dynamical equivalent of the pdf going to zero for $A=0$ \cite{RM10_1,BT11}. The approach and the results are in the same spirit if the coupling $g_2$ is taken into account.

\subsection{Computation of the autocorrelation function \label{Ansc}}

  We first study the time evolution of the modulus of the order parameter $A$ around the steady state $\langle A\rangle =\sqrt{\tilde{\epsilon}/g_1}$
for $\tilde{\epsilon}>0$. For that matter, we use the time autocorrelation function : $\langle f(t) f(t')\rangle$ with $f=A-\langle A\rangle$. This function contains information that simpler estimators does not. Firstly, the amplitude of this function indicates the corresponding amplitude of the time fluctuations of $A$, and the decay time of this function indicates the correlation time of $A$. Secondly, when examined in more details, the shape of this function indicates in which regime the transitional flow is. In this section we examine the case where relaminarisation and orientation fluctuations are so rare that one can consider that they do not occur. This will yield a specific type of correlation function which goes hand in hand with the parabolic shape of the large deviation function $I$ (Eq.~(\ref{LDP})) \cite{EPJB}. Note that near $R_{\rm g}$, where relaminarisations can occur and the correlation function, along with the large deviation function, is quite different \cite{cras,EPJB}.

\subsubsection{Analytical treatment}

The evolution of the order parameter around equilibrium is given by equation~\ref{eqchvar}.
One can neglect the drift term in the evolution equation, which now reads:
\begin{equation}\tau_0 \partial_t f =-e f +\zeta_2(t)\,,\end{equation}
  with $e=2\tilde{\epsilon}$, if the evolution is well described by the potential (Eq.~(\ref{eqchvar})). The expansion around equilibrium, with a decay rate $e$, is still valid, even when one enters the range of $R$ in which the deterministic part of the evolution equation is more complex than the Landau--Langevin equation \cite{RM10_1}.
The solution of the equation is:
\begin{equation} f(t)=\frac{1}{\tau_0}\int_0^t {\rm d}t' \exp\left(-\frac{e}{\tau_0}(t-t')\right)\zeta_2(t')\,.\label{solsc}\end{equation}

One has the product $f(t)f(t')$:
\begin{equation} f(t)f(t')=\int_0^t\int_0^{t'}{\rm d}t''{\rm d}t'''\left(\frac{1}{\tau_0^2}  \zeta_2(t'')\zeta_2(t''') \times\exp\left(-\frac{e}{\tau_0^2}\left(t+t'-t''-t''' \right) \right)\right)\,.\end{equation}
Taking the ensemble average, it yields:
\begin{equation} \langle f(t)f(t')\rangle=\int_0^t\int_0^{t'}{\rm d}t''{\rm d}t''' \left(\frac{1}{\tau_0^2} \alpha^2 \delta (t''-t''') \times\exp\left(-\frac{e}{\tau_0}\left(t+t'-t''-t''' \right) \right)\right)\,.\end{equation}
The time $t$ is taken smaller than $t'$ (the choice is of no consequence to the rest of the derivation):
\begin{equation}\notag\langle f(t)f(t')\rangle=\frac{\alpha^2}{\tau_0^2}\exp\left(-\frac{e}{\tau_0}(t+t') \right)  \int_0^t{\rm d}t'' \exp\left(-\frac{e}{\tau_0}\left(-2t'' \right) \right)\,,\end{equation}
\emph{i.e.}
\begin{equation}
\langle f(t) f(t')\rangle=\frac{\alpha^2}{2e\tau_0} \exp\left( -e\frac{(t+t')}{\tau_0}\right)\left(\exp\left(2e\frac{t'}{\tau_0}\right)-1\right)
\,.\label{auto}
\end{equation}
A few simplifications can be performed if $t=t'$:
\begin{equation}
\langle f(t) f(t)\rangle=\frac{\alpha^2}{2e\tau_0} \left(1-\exp\left(-\frac{2et}{\tau_0}\right)
\right)\,,\label{limapprox}
\end{equation}
and if $t$ and $t'$ are very large compared to $\tau_0/e$:
\begin{equation}
\langle f(t) f(t')\rangle=\frac{\alpha^2}{2e\tau_0} \exp\left( -\frac{e}{\tau_0}|t-t'|
\right)\,.\label{auto_}
\end{equation}
This yields classical exponentially decreasing time correlations.

\subsubsection{Numerical treatment of the model}

\begin{figure}\centerline{\includegraphics[width=6cm]{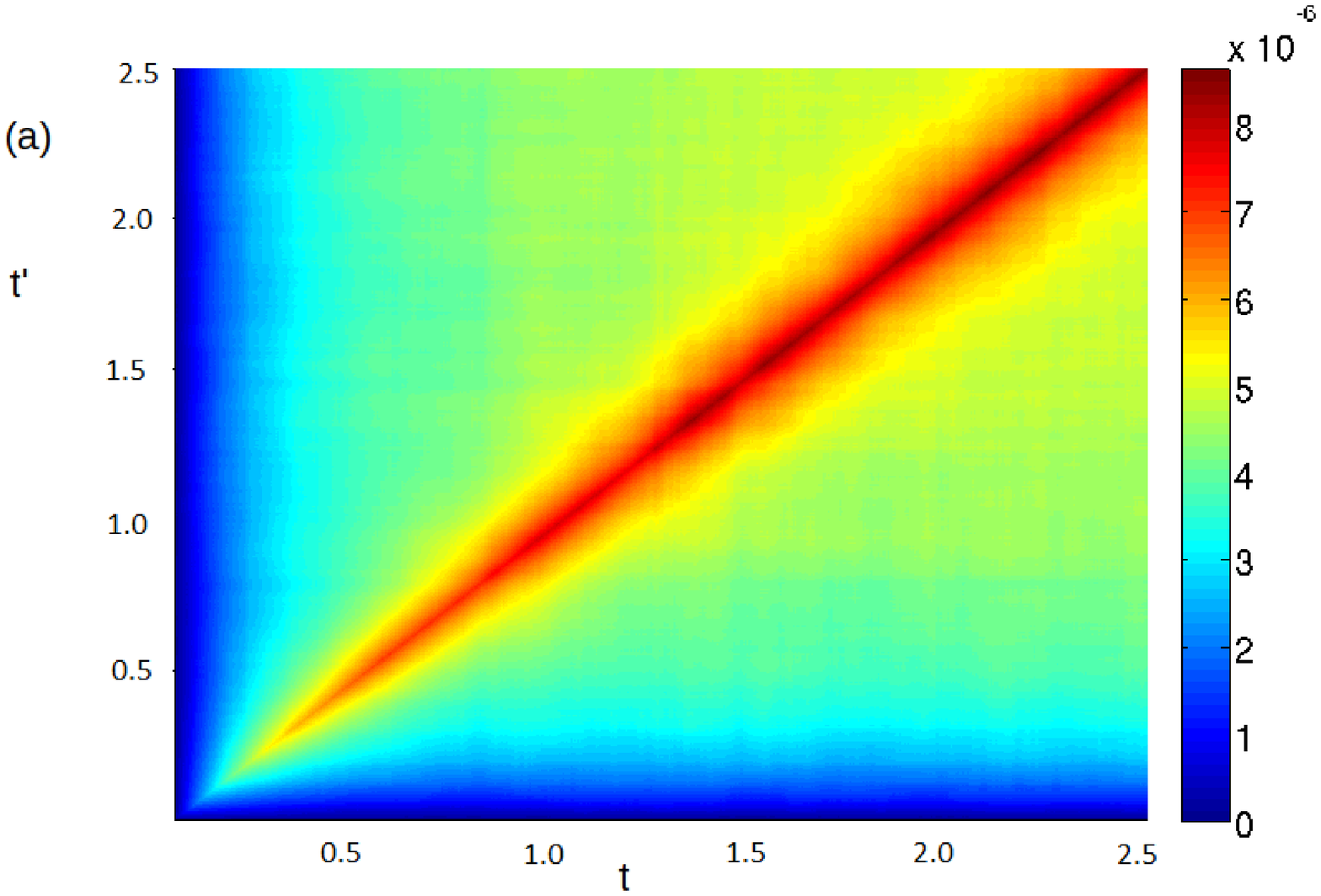}
\includegraphics[width=6cm]{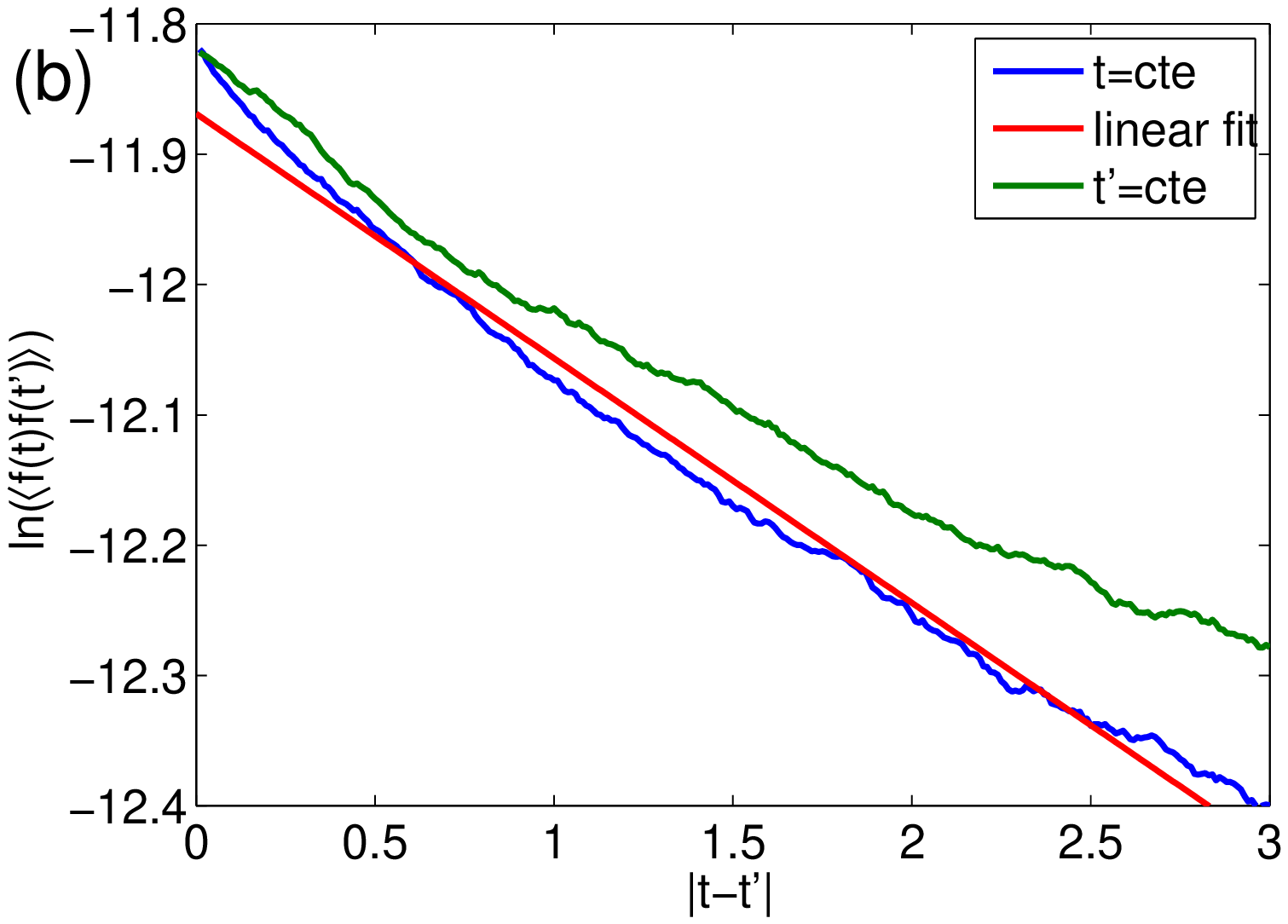}}\caption{Result of the time integration of the Landau--Langevin model : (a) autocorrelation function $\ln\langle f(t)f(t')\rangle$ in the plane $t$, $t'$, (b) : autocorrelation as a function of $|t-t'|$.}\label{fig38}\end{figure}

The autocorrelation function can be computed from ensemble averaging of the time series, divided in $N$ times series of duration $T_0$, in the same way as the fluctuations of the phase \cite{RM10_1}:
\begin{equation}\notag \langle f(t)f(t')\rangle =\frac{1}{N}\sum_{i=1}^N\left[
\left(f(t+(i-1)T_0)-f((i-1)T_0)\right) \left(f(t'+(i-1)T_0)-f((i-1)T_0)\right)\right]\,.\label{flucmod}
\end{equation}
The analytical result and the processing procedure can be validated by the numerical simulations of the Landau model. In a test case, for arbitrary values of the parameters ($\tau_0=1$, $\tilde{\epsilon}=0.29$, $\alpha=0.002$, $g_1=1$, $g_2=2$), we compute the autocorrelation in the $t,t'$ plane (Fig.~\ref{fig38} (a)). In figure~\ref{fig38} (b) we display the logarithm of the autocorrelation function as a function of $|t-t'|$ by varying $t$ (resp. $t'$) and keeping $t'$ constant (resp. $t$). The logarithm of the autocorrelation function along a $t$ or $t'$ constant line confirms that the prediction of an exponential decrease is good (Fig.~\ref{fig38} (b)). A fit of the logarithm by $-a|t-t'|+b$, yields $a=0.19$ and $b=-12$, to be compared respectively to $2\tilde{\epsilon}/\tau_0= 0.58$ and $ \ln(\alpha^2/(4\tilde{\epsilon}\tau_0))=-12.2$. The comparison to the value of $\alpha^2/{4\tilde{\epsilon}}\tau_0$ shows an error of only $15$\%. However, the error is much larger in the case of $2\tilde{\epsilon}/\tau_0$: although the laws are correctly predicted by analytics, there is an overestimate of the constant $\tau_0$. We will be wary of that fact in the analysis of the numerical simulation of the Navier--Stokes equation.

\subsection{The growth of the order parameter \label{s3T}}

\subsubsection{Principle}

  We now study the growth of the order parameter from the original noise. This models the appearance of the band from uniform turbulence. Our goal is
to propose a description of the growth that goes beyond the exponential growth and to describe the non linear effect and the effect of noise on the growth. In practice this amounts to the solving of equation~(\ref{eqchvar}):
\begin{equation}
\tau_0 \partial_t A=\tilde{\epsilon} A-g_1A^3+\frac{\alpha^2}{2A}+\zeta_2\,,
\end{equation}
starting from a nearly zero initial condition $A_0$ . We will use two strategies, both of them being of the stochastic structural stability theory type \cite{S3T}, and we will determine the proper ``nearly zero'' that describes uniform turbulence. These theories consist of computing the ensemble average trajectory $A(t)$ and the fluctuations around them, $\sigma(t)$. Most trajectories will be contained within $\pm\sqrt{\langle \sigma^2 \rangle}(t)$ of the average. This function indicates the shape that a group of independent realisations of the growth display when put together.

  The first approach consists of a  mean field approximation: the fluctuations, of the order of the noise $\alpha$, are assumed to be small relative
to the average. The ensemble average $A$ verifies the equation where the noise $\zeta$ is removed, with the initial condition $A_0$.
Meanwhile the fluctuations $\sigma(t)$ are treated as a linear perturbation of the deterministic part of the equation , of the
same order as the noise. It is somewhat artificial but the qualitative content appears more clearly.

    The second approach consists of solving in the limit of $\epsilon$ not too small compared to $\egb=\alpha^2g_1$  by using an
expansion in $\egb/\tilde{\epsilon}$:
\begin{equation}
A(t)=B_0(t)+\sqrt{\frac{\tilde{\epsilon}}{g_1}}\sum_{n=1}^{\infty}\left(\frac{\egb}{\tilde{\epsilon}}\right)^\frac{n}{2}B_n(t)\,,
\label{exp}
\end{equation}
with the initial conditions $B_{n>0}(t=0)=0$ and $B_0(t=0)=A_0$. Due to the expansion, each $B_{n>0}$ is the solution of a linear Langevin equation, with time varying coefficients. Since all of them are real, no resonance appears and solvability conditions are unnecessary. This expansion is more rigorous and robust in terms of order of approximation but less tractable. It mainly helps to validate the mean field strategy. Both solutions introduce the dimensionless parameter $\ggb\equiv2\alpha^2g_1/\tilde{\epsilon}^2$, a Ginzburg criterion type term \cite{LL}, which weighs the fluctuations $\alpha/\sqrt{\tilde{\epsilon}}$  (see \S~\ref{fluc} or \cite{RM10_1}) over the amplitude of the average $\sqrt{\epsilon/g_1}$. The mean field approach is valid and the expansion converges rapidly only if $\ggb$ is small.

In order to understand the features of the expanded solution, one can truncate the expansion at $n=2$ and compute $B_0$, $B_1$ and $B_2$. Indeed, at the first non-zero order in noise, the average $\langle A\rangle(t)$ only retains the terms $\langle B_0 \rangle+\alpha^2\langle B_2\rangle$, while the fluctuations are $\langle (A-\langle A\rangle)^2\rangle=\langle (B_0-\langle B_0\rangle)^2\rangle+\alpha^2\langle B_1^2\rangle$. This is because the first order contains $A_0$, which is decorrelated from $\zeta$, $B_1$ contains $\zeta$ and $B_2$ contains a deterministic term and $\zeta^2$. This order of approximation is a more rigorous formulation of the first method.

We proceed in the following way. We first compute the distribution of initial conditions. We then present the ensemble averages and fluctuations solutions. We eventually summarise the analysis and discuss the specificity of the evolution of the solutions.

\subsubsection{Distribution of initial conditions \label{fluc}}

  We first compute the initial condition $A_0$. We follow the mean field approximation of earlier studies of the bands \cite{RM10_1}.  In a former article \cite{RM10_1}, the average of the order parameter in the Landau--Langevin limit was obtained in the mean field limit by computing the minimum of the potential $V$ (Eq.~(\ref{eq12})). In the spirit of the mean field, the pdf $\rho$ is approximated by a Gaussian centered around the minimum of $V$:
\begin{equation}\rho\propto \exp(-(A-\langle A\rangle)^2/(2\gamma^2)\,.\label{ipdf}\end{equation}
If turbulence is uniform, one finds the minimum $\langle A\rangle =\alpha_</\sqrt{2|\tilde{\epsilon}_<}|$. We use $\alpha_<$ and $\tilde{\epsilon}_< <0$ so as to differentiate them from the value they take during the growth of the band $\tilde{\epsilon}>0$. In order to obtain the fluctuations $\gamma$, one simply has to expand $V$ to the second order and identify the corresponding coefficient with $\gamma$. One then finds: the fluctuations $\gamma= \alpha_</\sqrt{2|\tilde{\epsilon}_<|}$. In order to weigh the amplitude of the initial condition relatively to that of the final value, we introduce another dimensionless parameter $r\equiv A_0 \sqrt{g_1/\tilde{\epsilon}}$.

\subsubsection{Mean field solution}

Starting from equation:
\begin{equation}
\tau_0 \partial_t A=\underbrace{\tilde{\epsilon} A-g_1A^3+\frac{\alpha^2}{2A}}_{F(A)}+\zeta_2\,,
\end{equation}
we apply the mean field approach.

In order to obtain the average, one has to solve:
\begin{equation}
\tau_0 \partial_t A=\tilde{\epsilon} A-g_1A^3+\frac{\alpha^2}{2A}\,,
\label{eqmn1}
\end{equation}
with the initial condition $A(t=0)=A_0$. The full resolution is somewhat involved, we describe it in appendix~\ref{Afull}. The solution is given by equation~(\ref{sol_gen}).

If we consider the noise $\alpha$, $\alpha_<$, or the amplitude of the non linearity $g_1$ to be small but non zero, we can simultaneously take the limit $r\propto \ggb\rightarrow0$ and keep the first non zero order in equation~(\ref{solrw}). This limit corresponds to the first stage of the growth.
This gives the equation:
 \begin{equation}
\notag A(t)=A_0\exp\left(\frac{\tilde{\epsilon} t}{\tau_0}\sqrt{1+\ggb}\right) \sqrt{(1-\frac{\alpha^2}{2\tilde{\epsilon} A_0^2})+\frac{\alpha^2}{2\tilde{\epsilon} A_0^2}\exp\left(-\frac{2\tilde{\epsilon} t}{\tau_0}\sqrt{1+\ggb}\right)}\,.\label{A_temps}
\end{equation}
 This reveals an interesting effect: it shows an increase of the growth rate with the noise when $\tilde{\epsilon}$ is small, embodied by $\ggb$. It strengthens the observation that the noise in the numerical solution of the stochastic GL equation modifies the growth rate \cite{phD}. Besides, it contains the trace of the fluctuations at $t=0$, that decreases exponentially with a rate $2\tilde{\epsilon}\sqrt{1+\ggb}/\tau_0$: the second term in the square root should decrease before we have clear exponential growth.

If one takes $\ggb =0$ in equation~(\ref{solrw}),  we find the classical solution of the Landau equation :
 \begin{equation}
A(t)=\sqrt{\frac{\tilde{\epsilon}}{g_1}}\frac{1}{\sqrt{1+\left( \frac{\tilde{\epsilon}}{g_1A_0^2}-1\right)\exp{-\frac{2\tilde{\epsilon} t}{\tau_0}}}}\,,\,
A_0 \ll \sqrt{\frac{\tilde{\epsilon}}{g_1}}\rightarrow A(t)\simeq A_0\exp\left(\frac{\tilde{\epsilon} t}{\tau_0}\right)\,,\, A\approx \sqrt{\frac{\tilde{\epsilon}}{g_1}} \rightarrow A=\sqrt{\frac{\tilde{\epsilon}}{g_1}}\left(1-\frac{1}{2}\exp(-\frac{2\tilde{\epsilon} t}{\tau_0}) \right) \,. \label{A_temps_bis}
\end{equation}
The first approximation described the growth of $A$ away from $0$. The second describes the convergence toward $\sqrt{\tilde{\epsilon}/g_1}$,
this contains a simplified version of the non-linear effects of the third stage, \emph{i.e.} saturation, without the competition between orientations.
The average in the expansion contains the same physics as in the mean field approximation (\S~\ref{Aexp}, Eq.~(\ref{dampa})). It validates this approximation for $\ggb$ small.

 The fluctuations $\sigma (t)$ are obtained by solving:
\begin{equation}\tau_0 d_t \sigma=\left. \frac{dF}{dA}\right|_{A(t)}\sigma+\zeta\,.\label{eqmn2}\end{equation}
 One can write $\sigma=\lambda C_1$, with $\tau_0d_t C_1=dF/dA|_{A}C_1$ and $C_1(t=0)=1$ and $\lambda(t)$ verifying $C_1\tau_0d_t \lambda=\zeta$ and $\lambda(t=0)=0$. One simply finds:
\begin{equation} C_1(t)=\exp\left(\int_{t'=0}^t{\rm d}\,t'\frac{1}{\tau_0}\left. \frac{dF}{dA}\right|_{A(t')}\right)\,,\,\lambda(t)=\frac{1}{\tau_0}\int_{t'=0}^t{\rm d}t' \zeta(t')\exp\left(-\int_{t''=0}^{t'}{\rm d}t''\,\frac{1}{\tau_0}\left.\frac{dF}{dA}\right|_{A(t'')} \right)\,, \end{equation}
which yields $\sigma$:
\begin{equation}
\sigma(t)=\frac{1}{\tau_0}\int_{t'=0}^t{\rm d}t' \zeta(t')\exp\left(\int_{t''=t'}^t{\rm d}\,t''\left. \frac{dF}{dA}\right|_{A(t'')} \right)\,.
\label{eqb1}
\end{equation}
The fluctuations have a zero average around the mean at order $1$ in $\alpha$, even in logarithmic scale, since one has: \begin{equation}\ln(A+\sigma)=\ln(A)+\sigma/A+O(\alpha^2)\label{logsig}\,.\end{equation}
We compute the explicit expression of the ensemble average of $\sigma$ and $B_1$ in Appendix~\ref{EAS}. The calculation is the same in both cases.

\subsubsection{Time evolution of fluctuation type terms\label{ftt}}

The time dependence of $\sigma(t)$ (Eq.~(\ref{eqb1})) and its moments are not obvious. We perform a few approximations to make them more explicit. At small times, $A$ is small and $dF/dA(A(t))$ can be expanded. In the first order, it is equal to $\tilde{\epsilon}$. Inserting in equation~\ref{flucb1}, one has:
\begin{equation}
\langle\sigma^2\rangle=\frac{\alpha^2}{\tau_0^2}\int_{t'=0}^t{\rm d}t'\exp\left(\frac{2\tilde{\epsilon}(t-t')}{\tau_0} \right)\,,
\end{equation}
which can be rewritten and integrated:
\begin{equation}
\langle\sigma^2\rangle=\frac{\alpha^2\exp\left( \frac{2\tilde{\epsilon}t}{\tau_0}\right)}{\tau_0^2} \left[-\frac{\tau_0}{2\tilde{\epsilon}}\exp\left( \frac{-2\tilde{\epsilon}t'}{\tau_0}\right) \right]_{t'=0}^t\,.
\end{equation}
One eventually finds:
\begin{equation}
\langle\sigma^2\rangle=\frac{\alpha^2}{2\tilde{\epsilon}\tau_0}\left(\exp\left(\frac{2\tilde{\epsilon}t}{\tau_0} \right)-1 \right)\,.
\label{flucres}
\end{equation}
Note that this means that in the growth phase, the fluctuations of the logarithm of the order parameter $\sigma_{\ln}$ given by equation~(\ref{logsig}) are constant. Indeed:
\begin{equation}\sigma_{\ln}\equiv \sqrt{\langle \ln(A)^2-\langle \ln(A)\rangle^2\rangle}\simeq\sqrt{\langle \sigma^2\rangle}/A\label{sln}\end{equation}
is the ratio of two functions growing exponentially at the same rate.

The fluctuations $\sigma$ and $B_{1,2}$ have two successive regimes: expansion then retraction of perturbations induced by noise. From a different point of view, the first regime is controlled by linear growth and the second is controlled by non-linear saturations. The quantity:
\begin{equation}\int_{t''=t'}^t {\rm d}t''\exp(dF/dA|_{A(t)})\,,\end{equation}
and more particularly the sign of $dF/dA|_{A(t)}$, controls the regime in which the band is. This is exact in the case of the expansion and in good approximation in the mean-field case. Indeed, when considering equation~(\ref{eqb1}), one can see that the amplitude of the noise will be multiplied or divided, depending on the sign of the quantity in the exponential. The value of $t$ (and the corresponding $A(t)$ at which it changes signs can be easily estimated in order zero in noise (Eq.~(\ref{A_temps_bis})). Indeed, one has:
\begin{equation}  \left.\frac{dF}{dA}\right|_{A(t)}=\tilde{\epsilon}-2g_1A^2(t)=\tilde{\epsilon}\frac{\left( \frac{1}{r}-1\right)\exp\left(-\frac{2\tilde{\epsilon} t}{\tau_0} \right)-1}{1+\left(\frac{1}{r}-1\right)\exp\left(-\frac{2\tilde{\epsilon} t}{\tau_0} \right)}\,. \end{equation}
The numerator cancels out provided the initial condition is small enough ($r>1$) and does so at:
\begin{equation} t_{0}=\frac{\tau_0}{2\tilde{\epsilon}}\ln\left(\frac{1}{r}-1 \right) \label{tps}\,.\end{equation}
One then has $\langle A(t)\rangle=\sqrt{\tilde{\epsilon}/(2g_1)}$ at the first non-zero order in noise amplitude.

At long times $t>t_{0}$, one can have a qualitative discussion about the behaviour of fluctuations (\S~\ref{ftt}). The rate at which they grow decreases in time, the expansion becomes slower and slower until time reaches $t_0$ (Eq.~(\ref{tps})), at which time non-linear effects are felt and the analytics depart from the behaviour of the flow. Indeed, they do not contain the competition between orientations $A_\pm$. The analytical description of the switching is limited to the transition from one well to another (\emph{via} residency times \cite{RM10_2}). After said competition, both analytics and the flow agree again. In that last regime, one finds that the fluctuations decrease in time, and connect with the autocorrelation studied in the next section.

At longer times, the same type of expansion that has been done at small times can be done, and yields the results of the former section, that is, the autocorrelation function.

Again, the fluctuations in the expansion contain the same content as that of the mean field solution (Eq.~(\ref{flucb1})), and recover the exponential decay of the initial fluctuations (Eq.~(\ref{dampa})). Note that the time evolution of this term is deterministic (exponential decay), only its amplitude and its sign is random.

\section{Numerical Procedure \label{num}}

\subsection{Numerical Simulation of Navier--Stokes equations \label{chan}}

  We simulate the incompressible Navier--Stokes equations in the plane Couette configuration (Fig.~\ref{fig1} (b)). We use periodical boundary
conditions in the streamwise ($\mathbf{e}_x$) and spanwise  ($\mathbf{e}_z$) directions. The velocities are made dimensionless by $U$ and
the sizes by $h$ and durations by $h/U$, which leaves the Reynolds number $R=hU/\nu$, with $\nu$ the kinematic viscosity as the control parameter of the flow.
 The whole behaviour of the flow is set by $R$ together with the sizes $L_x$ and $L_z$ \cite{PM}.

  The numerical integration is performed using the code {\sc Channelflow} written by J. Gibson \cite{gibs}.
We follow two different procedures. On the one hand, we perform classical Direct Numerical Simulations (DNS), with a resolution of $8/3$ dealiased Fourier modes per unit length in the streamwise and spanwise directions and $27$ Chebychev modes in the wall normal direction. This number of modes has proven sufficient to obtain well resolved DNS \cite{MR10,dsc10}. On the other hand, we use the code to run a reduced order model. For that matter, we use $1$ dealiased Fourier mode per unit length in the streamwise direction, $2$ dealised Fourier modes per unit length in the spanwise direction and $15$ Chebychev modes in the wall normal direction. This approach gives reliable representations of the bands and does not alter the physics of the flow \cite{MR10,RM10_1}. The price to pay is a decrease of the transition thresholds from $[R_{\rm q}; R_{\rm t}]\approx [325 ; 415]$ to $[275 ; 355]$ \cite{MR10}.

  For both procedures, the oblique bands (Fig.~\ref{fig1} (a)) are obtained in the following way : we start from a smooth random initial condition.
It is then integrated in time at $R=500$ (for DNS) or at $R=450$ (low order modeling) for a duration of $500$. This gives a realistic uniformly
turbulent flow. It can be used as an initial condition for a quench : a sudden decrease of the Reynolds number (\S~\ref{gr}, see \cite{bddm} for laboratory experiments, or \cite{EPJB} for numerical experiments with our settings).
Further integration in time at said Reynolds numbers can be done in order to produce decorrelated initial conditions for quenches.

In order to obtain bands at $R\in [R_{\rm g}; R_{\rm t}]$, we start from one of the uniformly turbulent initial conditions, decrease the Reynolds number to $R$, then integrate for a duration of $1500$. The flow obtained can then be used as an initial condition for a simulation of the bands at $R$.

\subsection{Processing of numerical data \label{proc}}

In order to follow the time evolution of the modulation of turbulence in the flow, we need a quantity which indicates whether the turbulence
is organised in bands and which orientation the bands take. For that matter, we take advantage of the sinusoidal modulation of turbulence
of the band \cite{prigent02,BT11,RM10_1}. Once the fundamental mode of the bands $(k_x,\pm k_z)$ is identified, two similar strategies
can be followed to define a so-called order parameter.
The first one is adapted if the size of the system ($L_{x,z}$) is clearly larger than the wavelengths of the bands $(2\pi/k_{x,z})$. One can then compute two Hilbert transforms of the signal measured (in numerical studies, the velocity field, or in the experimental studies the light intensity) \cite{prigent02,phD}. This yields two complex functions $a_+(x,z,t)$ and $a_-(x,z,t)$ which vary more slowly in space than to the wavelength of the band. They are filtered. Their moduli give the amplitude of the modulation of the respective orientation at position $(x,z)$ at time $t$ and their phase give the relative shift of each patch of band.

 The second strategy is adapted if the size of the system is comparable to the wavelengths of the bands. Following Tuckerman \& Barkley \cite{BT11}, we proposed \cite{RM10_1} to use the Fourier transform of the $x$ component of the velocity field to define the order parameters $a_\pm=m_\pm(t)e^{\imath\phi_\pm(t)}$  by computing:
\begin{equation}
  m_\pm^2=\frac{1}{2}\int_{y=-1}^{y=1}|\hat{u}_x|^2(k_x,y,\pm k_z){\rm d}y \,,\,
  \phi_\pm=\arg\left(\hat{u}_x(k_x,0,\pm k_z\right)\,.
\end{equation}
The phases $\phi_\pm$ of the order parameters give the relative position of the respective orientation of the band in the domain, while the moduli of the order parameters give the amplitude of the modulation of the respective orientation. In this article, we follow this approach. Since we do not study the behaviour of the phases, we only use $m_\pm$, termed the order parameters by an abuse of language. This quantity is to be compared to the modulus of the order parameter $|A_\pm|$ studied theoretically in the former section. The order parameters are the ideal quantities to study the behaviour of the band itself: unlike the turbulent fraction, they capture the spatial organisation of turbulence.
\section{Numerical results \label{res}}

In this section, we present the results of the numerical simulations and compare them with theory. We follow the same approach as in the former theoretical section, firstly considering the evolution around the equilibrium position (\S~\ref{eq}) and secondly detailing the growth of the order parameter (\S~\ref{gr}).

\subsection{Around equilibrium \label{eq}}

  We first examine the fluctuation of the band around its equilibrium, using the time autocorrelation function. For that matter, we produce bands using our procedure. Since the time series required are extremely long, we use the reduce order procedure.  We sample a time series of the order parameter $m$ of duration $400000h/U$ at $R=290$, far from $\rt=355$. This Reynolds number corresponds to $\tilde{\epsilon}\simeq0.18$. This ensures that the orientation will not change. The size of the domain is $L_x\times L_z=110\times 64$, in order to accommodate one band.

\begin{figure}
\centerline{\includegraphics[width=6cm,clip]{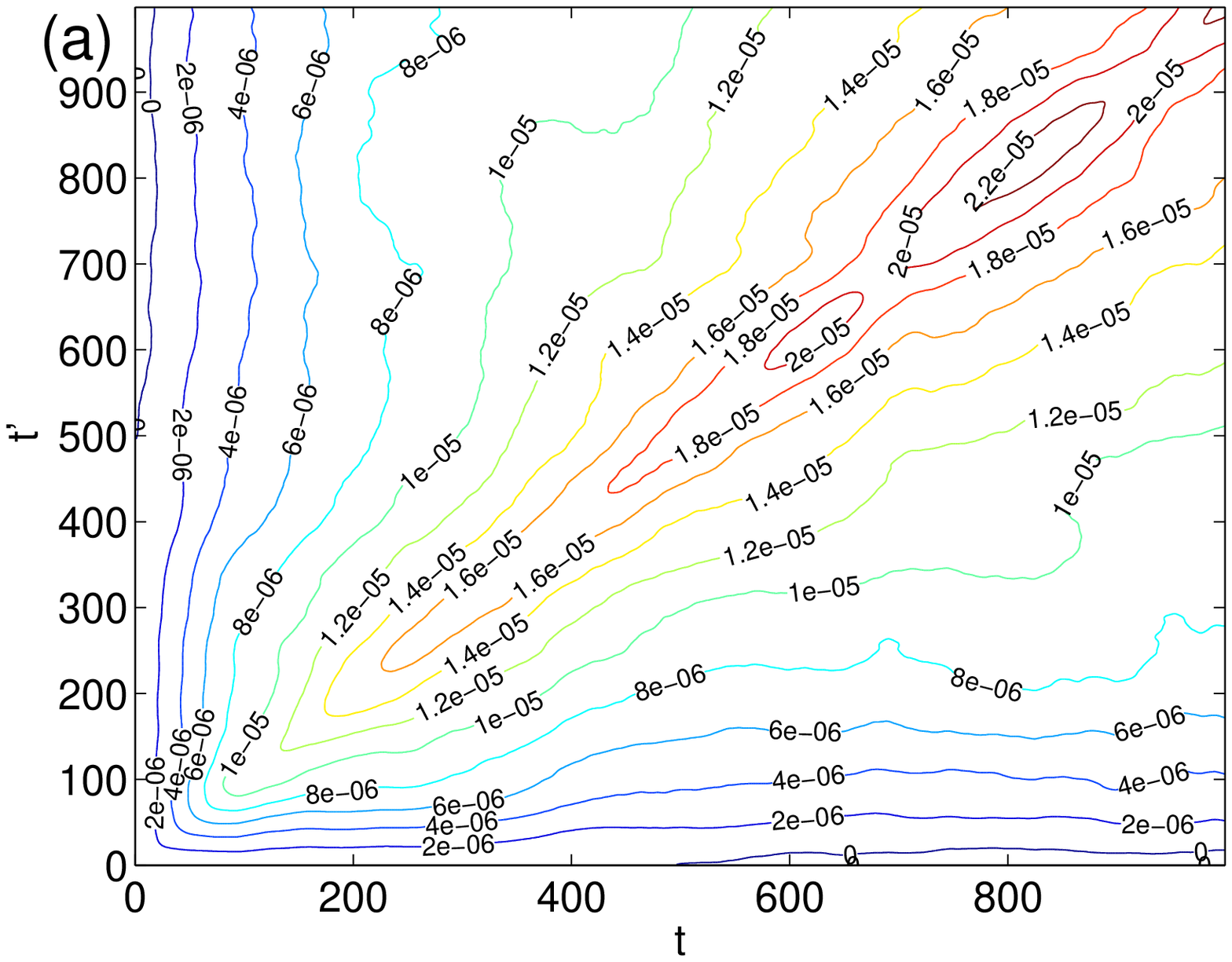}\includegraphics[width=6cm,clip]{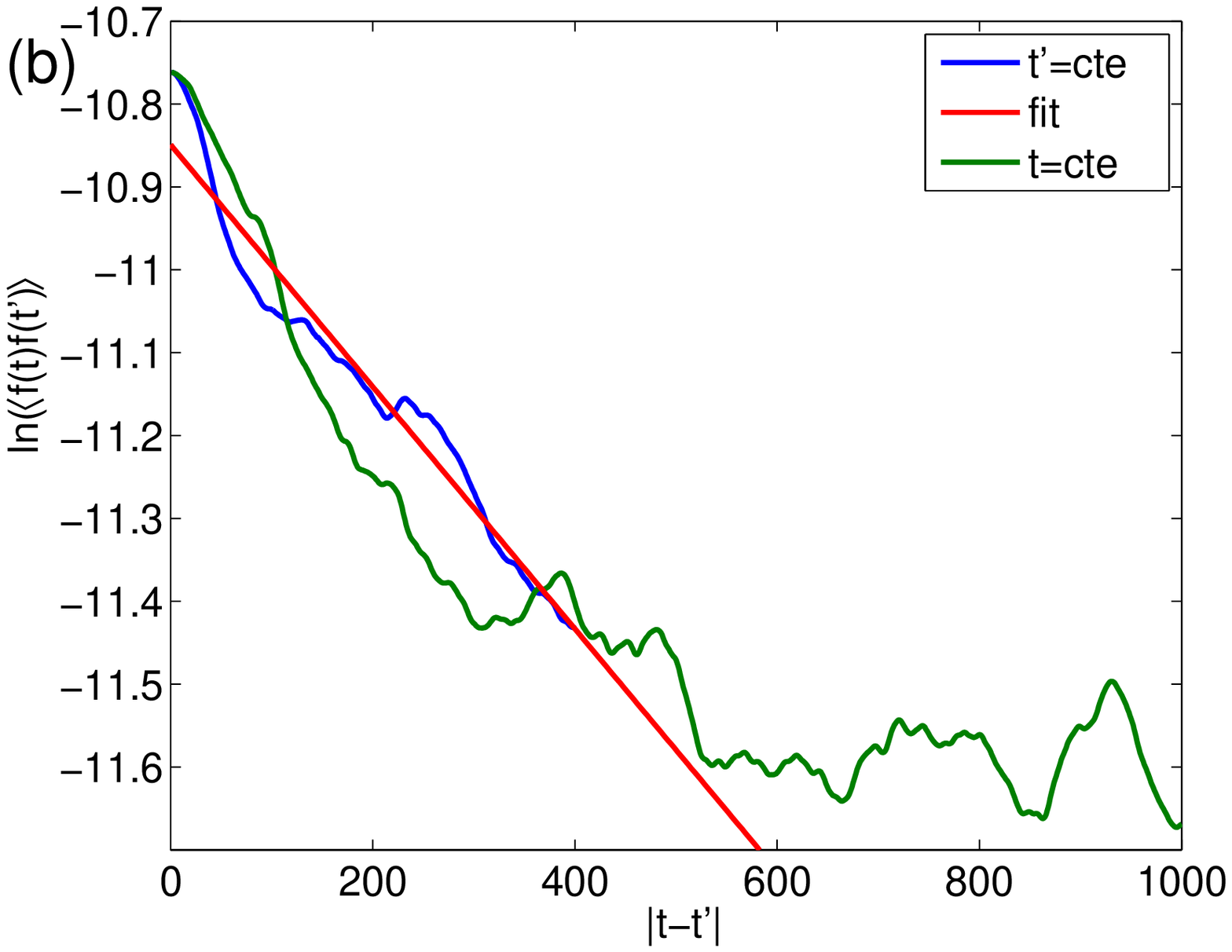}\includegraphics[width=6cm,clip]{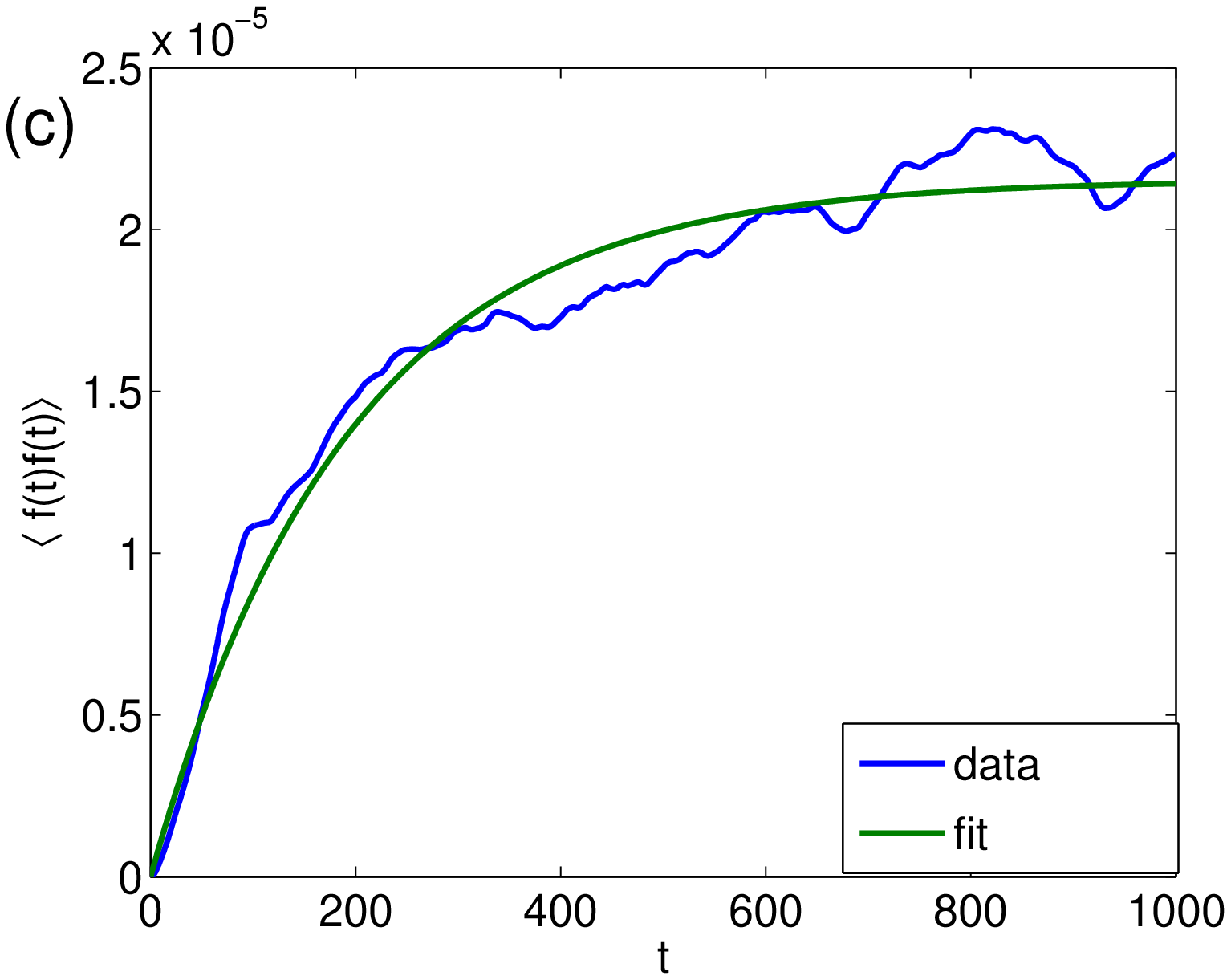}}
\caption{Processing of the time series of the order parameter extracted from numerical simulations of the Navier--Stokes equations at $R=290$, $N_y=15$ : (a) autocorrelation function in the $t,t'$ plane, (b) : logarithm of the autocorrelation function as a function of $|t-t'|$, (c) : autocorrelation function at $t=t'$.}\label{fig38_}
\end{figure}

  The autocorrelation function is computed from the time series using equation~\ref{flucmod}. The results are presented in figure~\ref{fig38_}. The
whole autocorrelation function is displayed in figure~\ref{fig38_} (a).  We display the logarithm of the autocorrelation function as a function of $|t-t'|$ in figure~\ref{fig38_} (b) by varying $t$ (resp. $t'$) and keeping $t'$ (resp. $t$) constant (in the same manner as Fig.~\ref{fig38} (b)). Moreover, we display the the correlation function as a function
of $t=t'$ in figure~\ref{fig38_} (c). The results are well fitted by the logarithm of equation~(\ref{auto_})
(linear decrease, Fig.~\ref{fig38_} (b)), and equation~(\ref{limapprox}) (exponential convergence, Fig.~\ref{fig38_} (c)). This indicates that the time evolution of the amplitude of modulation is very well described by the Ornstein-Uhlenbeck process derived from the GL model (Eq.~\ref{GLL}).

The fit yields the values  $\tau_0\simeq 68$ (if $t=t'$) and $\tau_0\simeq 257$ (function of $|t-t'|$) for the characteristic time
and $\alpha^2/(4\tilde{\epsilon}\tau_0)\simeq 2\cdot10^{-5}$, directly related to the amplitude of the noise felt by the bands. One finds discrepancies for $\tau_0$, which calls for an alternate estimation. Indeed, the study of the random walk of the phase yields $\alpha/\tau_0=4\cdot10^{-4}$ \cite{RM10_2}. In order to estimate the amplitude of the noise, a value of $\tau_0=30$ (found for counter rotating Taylor--Couette flow) was used, yielding $\alpha\simeq 0.012$. The results we have here allow for an estimation of $\alpha$ independently of $\tau_0$, in a situation where the validation of the procedure and the analytics by numerics showed no incertitude. This yields $\alpha\simeq0.036$. Knowing $\alpha$ allows for estimation of $\tau_0\simeq90$ independently of the uncertain fits.

\subsection{Growth \label{gr}}

\begin{figure}[!ht]\centerline{\includegraphics[width=7.5cm]{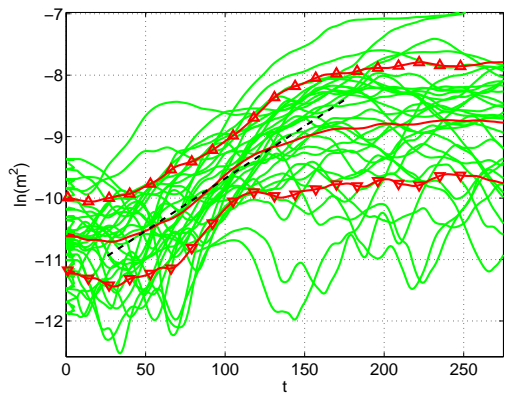}}
\caption{Logarithm of the square of the order parameter. The bright green lines corresponds to the time series. The bright red lines indicate the average. The bright red lines with upward and downward trianglesindicate the average plus/minus the fluctuations. The dashed red line indicates the linear fit of the data in the relevant range.} \label{trempe5}
\end{figure}

In order to study the appearance and organisation of the bands, we monitor the growth of the order parameter $m_\pm$ in DNS of quenches in a domain of size $L_x\times L_z=110\times 32$, from $R=500$ to $R=370$, with the high resolution. The uniformly turbulent initial condition is integrated up to $T=275$ for sixteen decorrelated initial conditions. Time series of the logarithm of the square of the order parameter are displayed together and discussed (Fig.~\ref{trempe5}). There are thirty two curves, since both orientations are concerned by the growth.

The amplitude of modulation $m_\pm$ starts from a small but non-zero value. In a first stage, it fluctuates around a constant value, as expected from the analysis of section~\ref{fluc}. This stage corresponds to the adjustment of the coherent structures to the new Reynolds number and the formation of laminar holes \cite{EPJB}. The amplitude of the modulation $m^2$ then grows randomly inside a flaring envelope in the second stage. However, in logarithmic scale, the time series of $m_\pm$ taken together reveal a well defined  bundle of curves with a constant thickness and a definite slope as noted from equations~(\ref{A_temps_bis}),~(\ref{logsig}) and~(\ref{flucres}) (Fig.~\ref{trempe5}). Eventually, the order parameters saturate, and a competition between each orientation takes place. This happens at a time $t$ which is approximately the half of the duration of the establishment of the bands, as noted from equation~(\ref{tps}). The competition is ultimately won by one of the two orientations, both of them having the same probability to arise. We compute the average of all the experiments $\langle m(t) \rangle$ as well as the standard deviation $\sigma_m(t)$ at each time. This gives us an estimate of the average evolution. In that case, the average is $\langle m(t) \rangle$ and the fluctuations around it are $\langle m(t) \rangle \pm \sigma_m$. These are indicated in logarithmic scale by the red lines in figure~\ref{trempe5}. Note that most time evolutions are contained within $\langle m\rangle \pm \sigma_m(t)$. These DNS results already show qualitative agreement with the S3T analysis of the growth (\S~\ref{s3T}). Indeed, the initial condition is non zero and distributed with a given variance. The order parameter first grows around an exponential trend with fluctuations of exponential amplitude around it. The logarithm of $m$ displays a linear tendency with constant fluctuations around it. The competition between orientation is not well predicted by analytics. Note that the growth is relatively fast, since $R=370$ is quite far away from $R_{\rm t}$ at this resolution.

  In order to perform a full quantitative comparison between the theory and DNS, we perform systematic quenches for a large range of $R$
with the low order procedure, since we need a very large number of repetitions. The simulations are performed in a domain of size $128\times 48$, which easily accommodate a band for all $R\in[R_{\rm g}; R_{\rm t}]$. For each Reynolds number, ten decorrelated initial conditions are quenched.

\begin{figure}
\centerline{\includegraphics[width=6cm,clip]{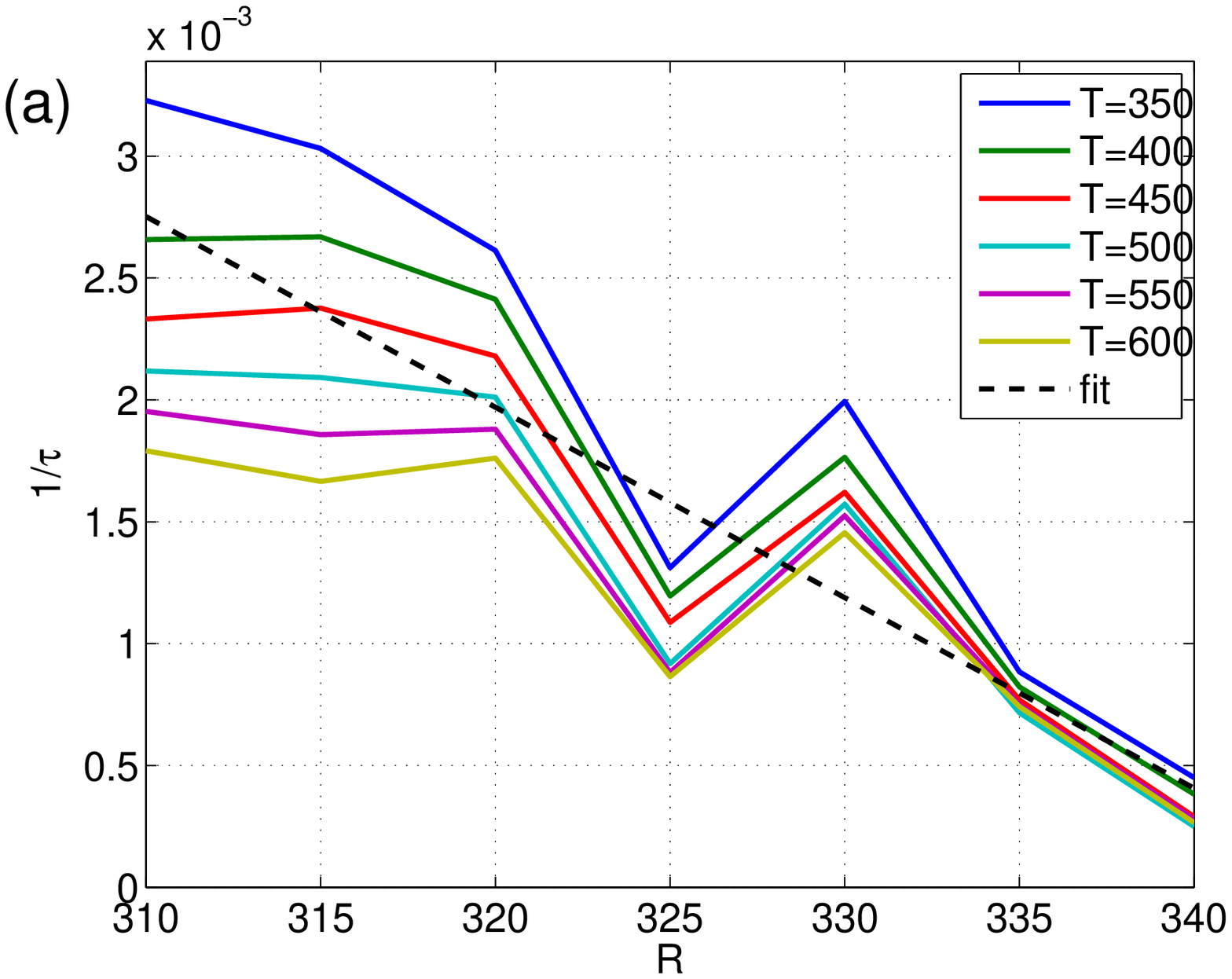}\includegraphics[width=6cm,clip]{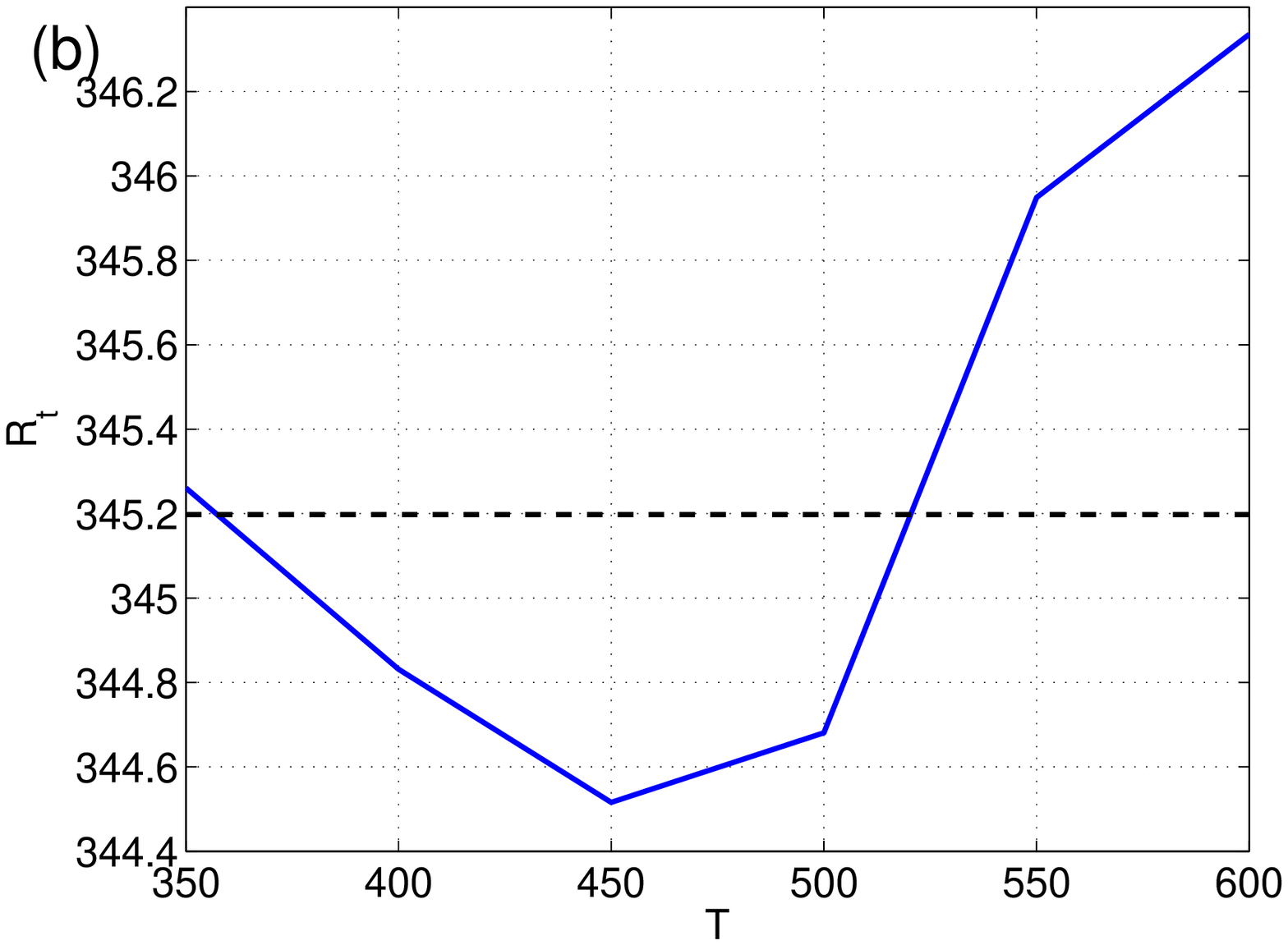}}
\caption{(a) : growth rate of the the order parameter, as a function of the Reynolds number ($N_y=15$) for several duration of fitting window. (b) : Threshold Reynolds number $R_{\rm t}$ as a function of the duration of the fitting window.}\label{taur}
\end{figure}

In order to extract the average behaviour of $m(t)$, the average $\langle m(t) \rangle$ is fitted in logarithmic scale in a window of duration $T$. This is motivated by the fact that the duration of the first exponential stage is not \emph{a priori} known. It should be bounded below, as noted from equation~(\ref{A_temps}). We found that starting the fit $100h/U$ after the beginning of the runs was a good compromise. It should be bounded above by $t_0$ (Eq.~\ref{tps}). In our study the fit do not use data past $700h/U$ after the beginning of the run. This assumes that the fluctuations $\sigma(t)$ around the mean will be averaged out, as predicted by the model. The growth rate is displayed as a function of the Reynolds number, for a range of fitting window duration $T$ in figure~\ref{taur} (a). Note that the characteristic time of organisation of the laminar-turbulent coexistence into bands measured here is several times larger than the characteristic time of formation of laminar holes measured from earlier numerical experiments \cite{EPJB}. The former are of order $O(10^2)$ while the later are of order $O(10^3)$.

For each value of $T$, $1/\tau$ is linear provided $R\gtrsim 320$. However, at constant $R$, the inverse decay shows large fluctuations around an average behaviour. A large incertitude is found on the growth rate, as was the case of Taylor--Couette flow \cite{phD}. Following the processing of the model, we perform a fit of all the $1/\tau$ lines in the range $320 \le R \le 340$ by $(1/\tau_0)(1-R_1/\rt)$ so as to minimise the incertitude on the fitting parameters. This gives $\rt \simeq345$ and $\tau_0\simeq 37$ (black dashed lines). One can alternatively fit the results of figure~\ref{taur} (a) line by line, for each fitting window. The results for $R_{\rm t}$, as a function of the fitting window duration, are displayed in figure~\ref{taur} (b).

There is a large incertitude on the characteristic time $\tau_0$. Indeed, the fitting parameter is $1/\tau_0$, as a consequence, the incertitude on $\tau_0$ is $\pm\tau_0^2 \Delta (1/\tau_0)\simeq 13$. The time scales $\tau_0$ computed here are consistent with the most precise estimation of the former section, given the error bars. Most fitting approach will only give a range for $\tau_0$, hence a large incertitude on the value of that parameter.

\begin{figure}
\centerline{\includegraphics[width=8cm]{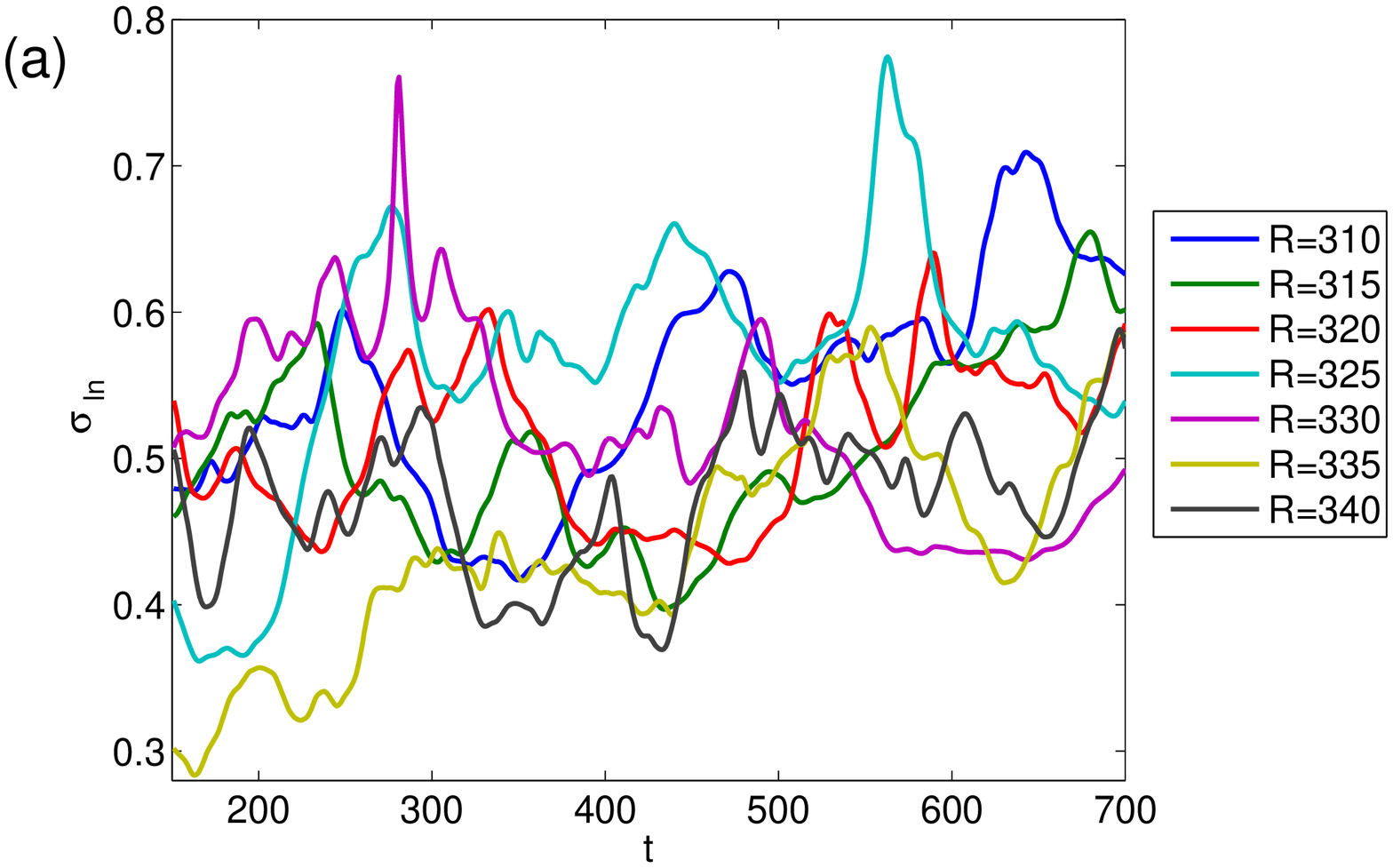}\includegraphics[width=6.5cm]{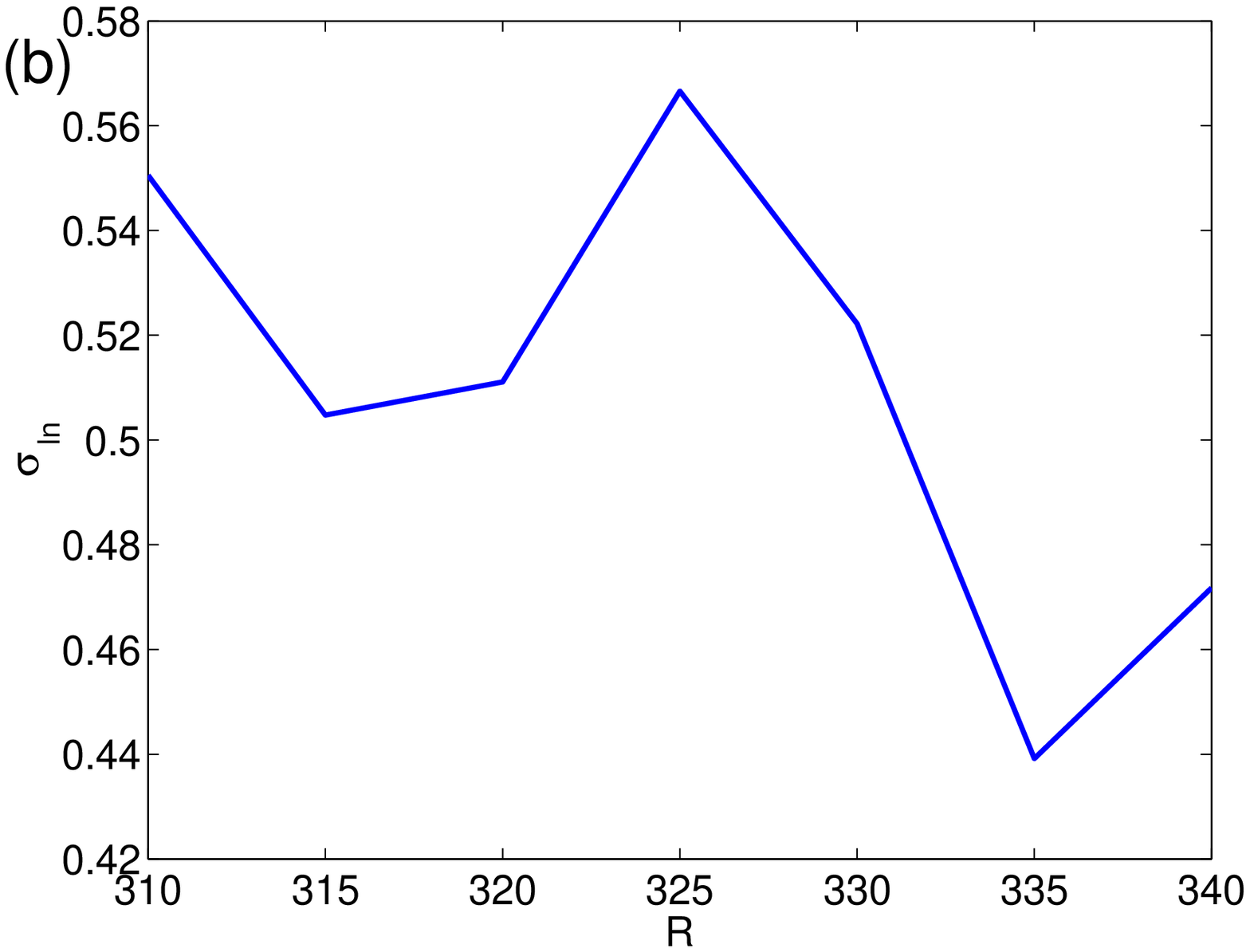}}
\caption{(a) : Time evolution of the fluctuations of the logarithm $\sigma_{\ln}$ around the average for $R\in [310 ; 340]$, in low order simulations. (b) time average of the fluctuations of the logarithm $\sigma_{\ln}$ around the average as a function of the Reynolds number.}
\label{figsig}
\end{figure}

We then compute systematically the fluctuations of the logarithm of $m$, $\sigma_{\ln}$  (Eq.~(\ref{sln})) at small time, in order to quantify the impression of figure~\ref{trempe5} (a,b). We use the same set of numerical experiments and compute the fluctuations in the same manner. The fluctuations as a function of time are displayed in figure~\ref{figsig} (a). In agreement with the analytical results, we find that $\sqrt{\langle\sigma^2\rangle}$ is a nearly constant quantity. However, it does not seem to depend on the Reynolds number. This assertion is verified in the case of the time average (Fig.~\ref{figsig} (b)) which shows no clear tendency. Overall, we find a value of approximately $0.5$, which should be compared to $\alpha\sqrt{g_1}/(\tilde{\epsilon}\sqrt{2\tau_0})$. Once known quantities $\tilde{\epsilon}\simeq 0.1$ and $g_1\simeq 100$ are removed, this gives us $\alpha^2/(\tau_0)=O(10^{-5})$. This shows reasonable quantitative agreement with the former section.

\section{Conclusion \label{conc}}

This article examined the stochastic evolution of the oblique modulation of turbulence in plane Couette flow.
We performed analytical and numerical treatment of the Stochastic Landau model that describes the bands in order to predict the behaviour of the flow
around its equilibrium as well as the growth of bands from uniform turbulence. We then validated this model against Direct Numerical Simulations. These results also explained observations on previous numerical simulations of the model \cite{phD}, such as the shift of the growth rate of the order parameter by noise.

The consequences of these results are twofold. On the one hand, we can consider quenches and the formation of laminar holes and then laminar turbulent bands from uniform wall turbulence. The comparison between the current results and those of a former article shows the clear timescale separation between the formation of laminar holes and the organisation into oblique bands \cite{EPJB}. The former phenomenon is most likely a local process, which is caused by viscous decay and failure of sustainment of turbulence. It should be present in all shear flows, including those which do not have a spanwise extension like Hagen--Poiseuille pipe flow. Meanwhile, the second phenomenon is most likely a combination of global processes, which involves the advection of small scale structure by large scale flows and feed back on turbulence (see \cite{PRL,ispspot,BT07} and references within).

On the other hand, these results further strengthen the assertion that the amplitude of modulation of turbulence is very well described by a Ginzburg--Landau equation with an additive noise. This means that some interesting properties of this type of equation should be expected for the modulation of turbulence. In particular, this allows us to easily predict facts that should manifest themselves in very large size systems. For instance, quenches in very large systems first lead to a spatial coexistence of domains of both orientations. It is very likely that the dynamics of the domain sizes is predicted by the GL equation, which means that we would find coarsening dynamics and specific exponents for the domain sizes as function of time (see \cite{EPJB,bray}). Moreover, it is also very likely that the behaviour predicted near $R_{\rm t}$: a critical phenomenon and the divergence of the response function of $m_\pm$ with size as $R\rightarrow R_{\rm t}$ should arise from the very intermittent regime of $R\simeq R_{\rm t}$. Early simulations indicate that this is very likely the case \cite{RNL}.

All things considered, the results of this article confirm earlier assumptions on both the sustainment mechanisms of the oblique bands and their large scale behaviour. This motivates further studies in both directions.

\section*{acknowledgement}

The author thanks the hospitality of the INLN (Universit\'e Nice Sophia Antipolis, France), where parts of this work where performed. He also thanks the GRADE language service of Goethe Univerist\"at, Frankfurt, for helpful comments and corrections on the manuscript. This work was granted access to the HPC and visualization resources of ``Centre de Calcul Interactif'' hosted by ``Universit\'e Nice Sophia Antipolis''.

\appendix

\section{Full solution for the mean field approach \label{Afull}}

In order to solve equation~(\ref{eqmn1}), we use a separation of variables, which gives:
\begin{equation}\int_{A_0}^{A(t)}\frac{AdA}{\left(\frac{\alpha^2}{2}+\tilde{\epsilon} A^2-g_1A^4 \right)}=\frac{t}{\tau_0} \end{equation}
The fraction in the integral can be rewritten as:
\begin{equation}
\frac{1}{2\tilde{\epsilon}\sqrt{1+\frac{2\alpha^2g_1}{\tilde{\epsilon}^2}}}\left(\frac{1}{A+\sqrt{\frac{\frac{\tilde{\epsilon}}{g_1}+\sqrt{\frac{\tilde{\epsilon}^2}{g_1^2}+2\frac{\alpha^2}{g_1}}}{2}}} +\frac{1}{A-\sqrt{\frac{\frac{\tilde{\epsilon}}{g_1}+\sqrt{\frac{\tilde{\epsilon}^2}{g_1^2}+2\frac{\alpha^2}{g_1}}}{2}}}  -\frac{1}{A+\sqrt{\frac{\frac{\tilde{\epsilon}}{g_1}-\sqrt{\frac{\tilde{\epsilon}^2}{g_1^2}+2\frac{\alpha^2}{g_1}}}{2}}}
-\frac{1}{A-\sqrt{\frac{\frac{\tilde{\epsilon}}{g_1}-\sqrt{\frac{\tilde{\epsilon}^2}{g_1^2}+2\frac{\alpha^2}{g_1}}}{2}}}
\right)\end{equation}
The integral is then straightforward and leads to:
\begin{equation}
\ln\left(\left(\frac{A^2(t)-\frac{\frac{\tilde{\epsilon}}{g_1}+\sqrt{\frac{\tilde{\epsilon}^2}{g_1^2}+2\frac{\alpha^2}{g_1}}}{2}}{A^2_0-\frac{\frac{\tilde{\epsilon}}{g_1}+\sqrt{\frac{\tilde{\epsilon}^2}{g_1^2}+2\frac{\alpha^2}{g_1}}}{2}}\right) \left(\frac{A^2_0-\frac{\frac{\tilde{\epsilon}}{g_1}-\sqrt{\frac{\tilde{\epsilon}^2}{g_1^2}+2\frac{\alpha^2}{g_1}}}{2}}{A^2(t)-\frac{\frac{\tilde{\epsilon}}{g_1}-\sqrt{\frac{\tilde{\epsilon}^2}{g_1^2}+2\frac{\alpha^2}{g_1}}}{2}} \right)\right)=\frac{2t\tilde{\epsilon}}{\tau_0}\sqrt{1+\frac{2\alpha^2g_1}{\tilde{\epsilon}^2}}
\end{equation}
 Note that $r$ it is of the same order of $\ggb$, $r\simeq \alpha_<\sqrt{g_1 \tilde{\epsilon}_</\tilde{\epsilon}}$.
A reorganisation then leads to the solution $A(t)$. The solution then reads:
\begin{equation}
A(t)=\sqrt{\frac{\tilde{\epsilon}}{g_1}}\sqrt{\frac{\frac{1+\sqrt{1+\ggb}}{2}\frac{r^2-\frac{1-\sqrt{1+\ggb}}{2}}{r^2-\frac{1+\sqrt{1+\ggb}}{2}}-\exp\left(-\frac{2t\tilde{\epsilon}}{\tau_0}\sqrt{1+\ggb}\right)\left( \frac{1-\sqrt{1+\ggb}}{2}\right)}{\frac{r^2-\frac{1-\sqrt{1+\ggb}}{2}}{r^2-\frac{1+\sqrt{1+\ggb}}{2}}-\exp\left(-\frac{2t\tilde{\epsilon}}{\tau_0}\sqrt{1+\ggb}\right)}}\,.\label{sol_gen}
\end{equation}

In order to examine the first stages of the growth of the order parameter, one can first rewrite this solution:
\begin{equation}
A(t)=A_0\exp\left(\frac{t\tilde{\epsilon}}{\tau_0}\sqrt{1+\ggb}\right)
\sqrt{\frac{\left(\frac{1+\sqrt{1+\ggb}}{2}-\frac{\ggb}{4r^2} \right)-\exp\left(-\frac{2t\tilde{\epsilon}}{\tau_0}\sqrt{1+\ggb}\right)\left(\frac{1-\sqrt{1+\ggb}}{2}-\frac{\ggb}{4r^2} \right)}{\left( r^2-\frac{1-\sqrt{1+\ggb}}{2} \right)\exp\left(\frac{2t\tilde{\epsilon}}{\tau_0}\sqrt{1+\ggb}\right)-\left(r^2-\frac{1+\sqrt{1+\ggb}}{2} \right)}}
\label{solrw}
\end{equation}

\section{Expansion up to order 2 \label{Aexp}}

In this appendix, we present the expansion of the solution up to order, in order to compare it to the mean field solution. The purpose is to validate the mean field approach.

\subsection{Expanded solution}

 In order to obtain the $0$th order of the expansion~(\ref{exp}), one has to solve:
\begin{equation}
\tau_0\partial_t B_0=\tilde{\epsilon} B_0-g_1B_0^3=\tilde{F}(B_0)\,,
\end{equation}
with the initial condition $B_0(t=0)=A_0$. The solution is simply~(\ref{A_temps_bis}).

 The first order of the expansion is then solution of:
\begin{equation}\tau_0 d_t B_1=\left. \frac{d\tilde{F}}{dA}\right|_{B_0(t)}B_1+\zeta^1\,,\end{equation}
with $\zeta^1$ of variance $1$ (rescaled by $\alpha$), and it is obtained in the same manner as the fluctuations in the mean field approach. One has:
\begin{equation}
B_1(t)=\int_{t'=0}^t{\rm d}t'\frac{1}{\tau_0} \zeta^1(t')\exp\left(\int_{t''=t'}^t{\rm d}\,t''\frac{1}{\tau_0}\left. \frac{d\tilde{F}}{dA}\right|_{B_0(t'')} \right)\,.
\label{eqb1_}
\end{equation}
Note that it has the same structure as $\sigma$.

  One can then obtain $B_2$ in the same fashion. It is solution of:
\begin{equation} \tau_0d_tB_2=\left.\frac{d\tilde{F}}{dA}\right|_{B_0}B_2+\underbrace{\frac{2\sqrt{\tilde{\epsilon}}}{B_0\sqrt{g_1}}+\frac{1}{2}\sqrt{\frac{\tilde{\epsilon}}{g_1}}\left.\frac{d^2\tilde{F}}{dA^2}\right|_{B_0}B_1^2}_{\mu}\,. \end{equation}
The term $2\sqrt{\tilde{\epsilon}}/(B_0\sqrt{g_1})$ comes from the drift term.
The function $\mu$ is random and contains $\frac{2\alpha^2}{B_0}$ and $\zeta^1(t')\zeta^1(t'')$. It has a nonzero average. One therefore has:
\begin{equation}
B_2(t)=\frac{1}{\tau_0}\int_{t'=0}^t{\rm d}t' \mu(t')\exp\left(\int_{t''=t'}^t{\rm d}\,t''\frac{1}{\tau_0}\left. \frac{d\tilde{F}}{dA}\right|_{B_0(t'')} \right)\,.
\label{eqb2}
\end{equation}

\subsection{Ensemble averages of solutions \label{EAS}}
Taking the averages, one finds $\langle B_1\rangle=0$. Besides:
\begin{equation} \langle B_2 \rangle= \frac{1}{\tau_0}\int_{t'=0}^t{\rm d}t'\exp\left(\frac{1}{\tau_0}\int_{t''=t'}^t{\rm d}\,t''\left. \frac{d\tilde{F}}{dA}\right|_{B_0(t'')} \right)  \left( \frac{2\sqrt{\tilde{\epsilon}}}{B_0\sqrt{g_1}} +\frac{1}{2}\sqrt{\frac{\tilde{\epsilon}}{g_1}}\left.\frac{d^2\tilde{F}}{dA^2}\right|_{B_0}\int_{t''=0}^{t'}{\rm d}t''\exp\left(\frac{2}{\tau_0}\int_{t'''=t''}^{t'}{\rm d}t''' \left.\frac{d\tilde{F}}{dA}\right|_{B_0(t'')} \right)\right)\label{flucb2}\end{equation}

 Moving to the fluctuations $\alpha^2\langle B_1^2 \rangle\simeq\langle \sigma^2\rangle$:
\begin{equation}\langle \sigma^2 \rangle=\frac{\alpha^2}{\tau_0^2}\int_{t'=0}^{t}{\rm d}t'\exp\left(\frac{2}{\tau_0}\int_{t''=t'}^{t}{\rm d}t'' \left.\frac{dF}{dA}\right|_{A(t'')} \right)
\label{flucb1}\end{equation}
We can neglect the difference between $\tilde{F}$ and $F$.
On can analyse the behaviour of this solution. It is $0$ at $t=0$ and has an exponential growth led by $dF/dA$ initially positive. When $dF/dA$ decreases, this growth is more limited, and if crosses $0$, the fluctuations then retract.

 The effect of the distribution of initial conditions on the $0^{th}$ and in $A$ in the limit $r\rightarrow 0$ order of the expansion can be shown by writing $A_0=\langle A_0\rangle+\delta A_0$. The departure $\delta A_0$ is distributed according to the initial pdf (Eq.~\ref{ipdf}). This yields after expansion of $B_0$:
\begin{equation}\notag B_0=\underbrace{\sqrt{\frac{\tilde{\epsilon}}{g_1}}\frac{1}{\sqrt{1+\left( \frac{\tilde{\epsilon}}{g_1\langle A_0\rangle^2}-1\right)\exp{-\frac{2\tilde{\epsilon} t}{\tau_0}}}}}_{\langle B_0\rangle} \left( 1+\underbrace{\frac{\left(\frac{\tilde{\epsilon}}{g_1} \right)}{1+\left( \frac{\tilde{\epsilon}}{g_1\langle A_0\rangle^2}-1\right)\exp{-\frac{2\tilde{\epsilon} t}{\tau_0}}}\delta A_0\exp\left(-\frac{2\tilde{\epsilon} t}{\tau} \right)+\ldots}_{\langle .^2\rangle\rightarrow\sigma_{B_0}} \right)\label{dampa}\end{equation}

That correction term stays small: it is well approximated by $\langle A_0\rangle^2g_1/\tilde{\epsilon}$ at $t=0$. One can note that the additional corrections all decrease exponentially with time.
The ensemble average of the expansion in $\sqrt{\egb/\epsilon}$ (Eq.~\ref{dampa} and particularly Eq.~\ref{flucb2}) has most of the content of the mean field solution. It contains the limit $r\rightarrow0$ only through $\langle B_2\rangle$. This term grows until non linearity $g_1$ is felt (Eq.~(\ref{tps})) and then decays. The whole expansion of $\langle B_{2p}\rangle$ terms may contain the modification of the growth rate, again of order $\ggb$.

\newpage
\thispagestyle{empty}
\begin{center}
{\Large \textbf{Graphical Abstract accompanying the article on the EPJE webside}}
\end{center}
\begin{flushleft}
\begin{pspicture}(21,12)
\rput(10,11.5){Amplitude of modulation for 16 trajectories}
\rput(10,9){\includegraphics[width=6cm]{m2_log_32_moy_fluc_fit.eps}}
\psline{->}(7.85,7)(2,5.68)
\psline{->}(8.6,7)(6,5.68)
\psline{->}(9.45,7)(10,5.68)
\psline{->}(10.3,7)(14,5.68)
\psline{->}(11.15,7)(18,5.68)
\rput(2,5){\includegraphics[width=4cm]{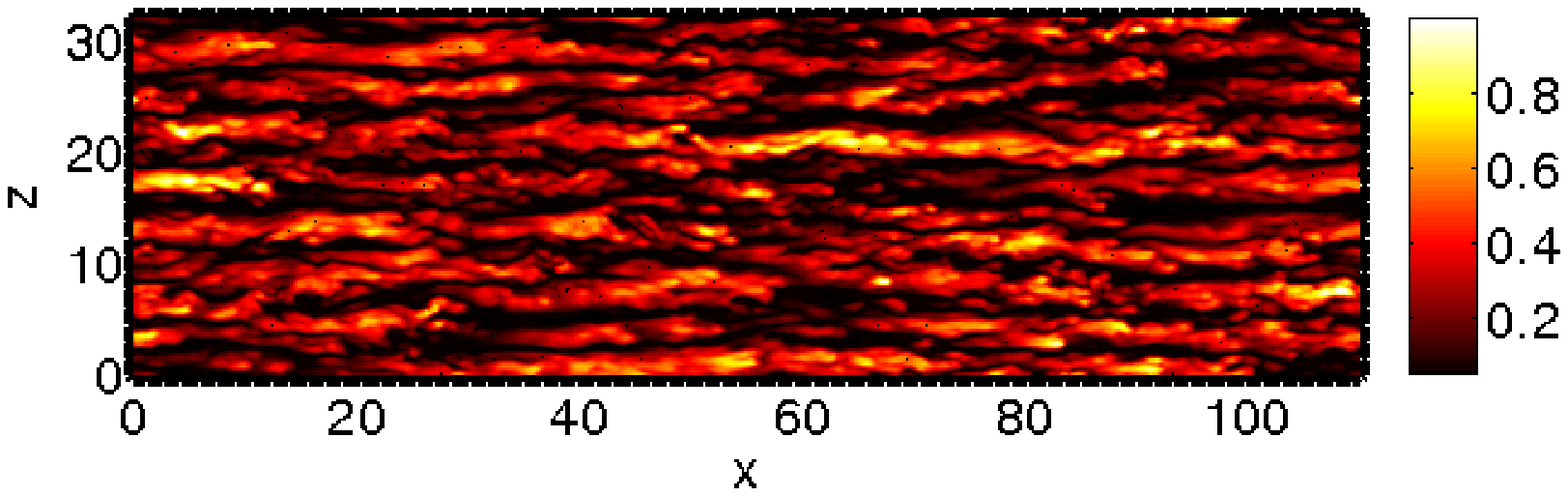}}\rput(6,5){\includegraphics[width=4cm]{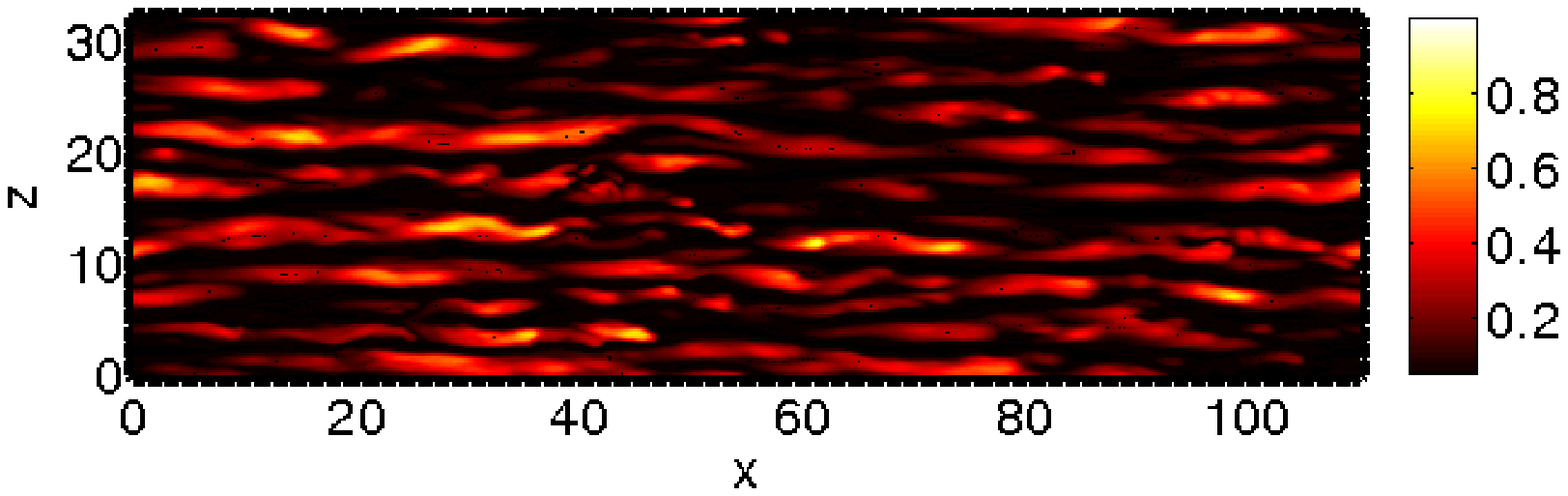}}
\rput(10,5){\includegraphics[width=4cm]{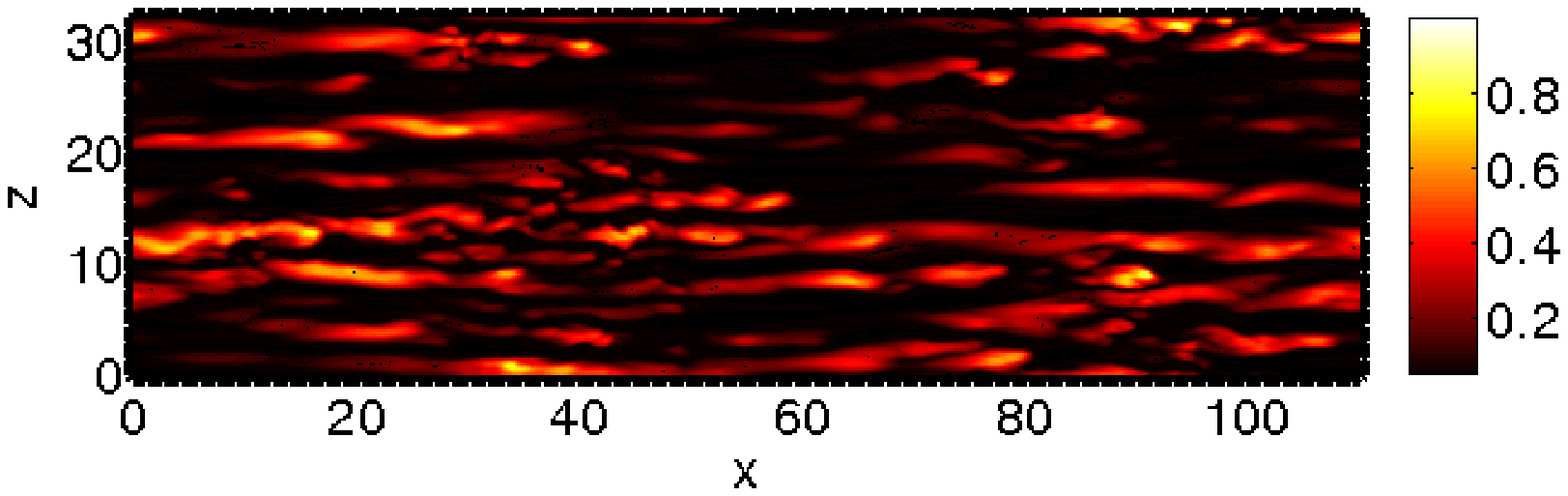}}\rput(14,5){\includegraphics[width=4cm]{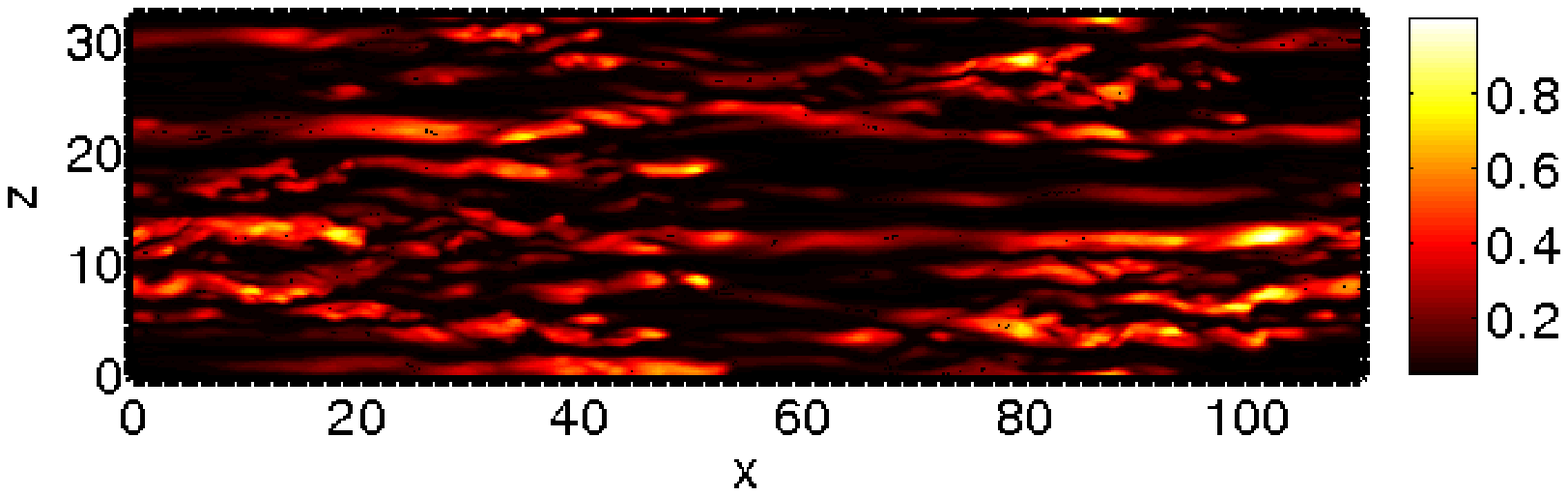}}
\rput(18,5){\includegraphics[width=4cm]{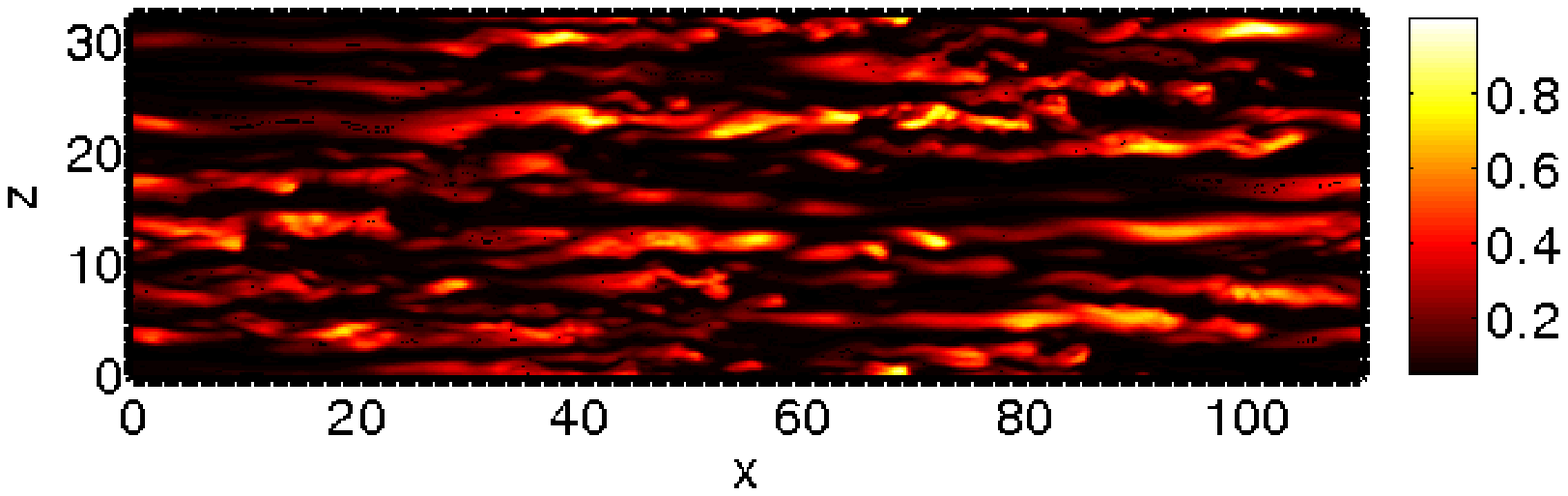}}
\rput(2,3.5){\includegraphics[width=4cm]{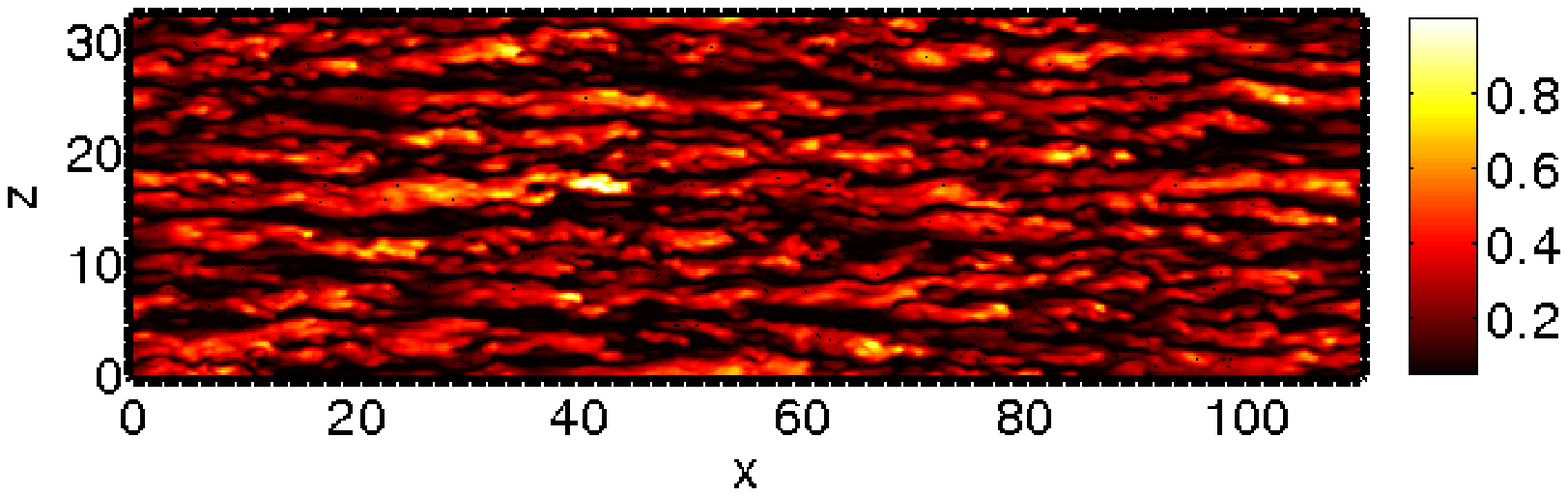}}\rput(6,3.5){\includegraphics[width=4cm]{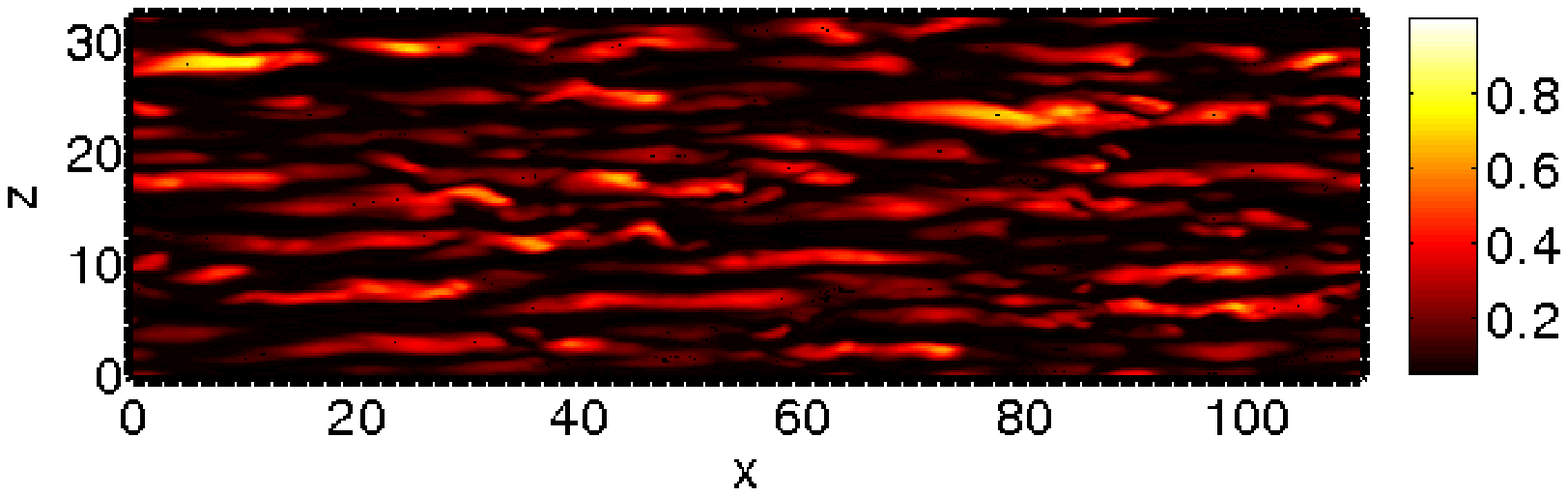}}
\rput(10,3.5){\includegraphics[width=4cm]{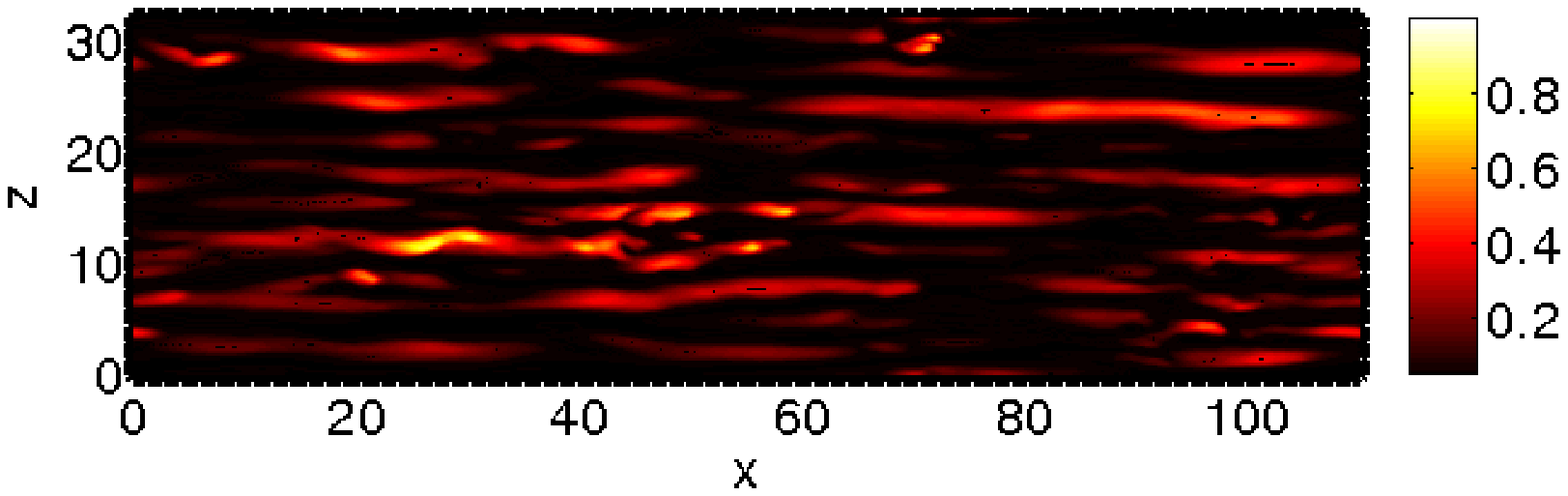}}\rput(14,3.5){\includegraphics[width=4cm]{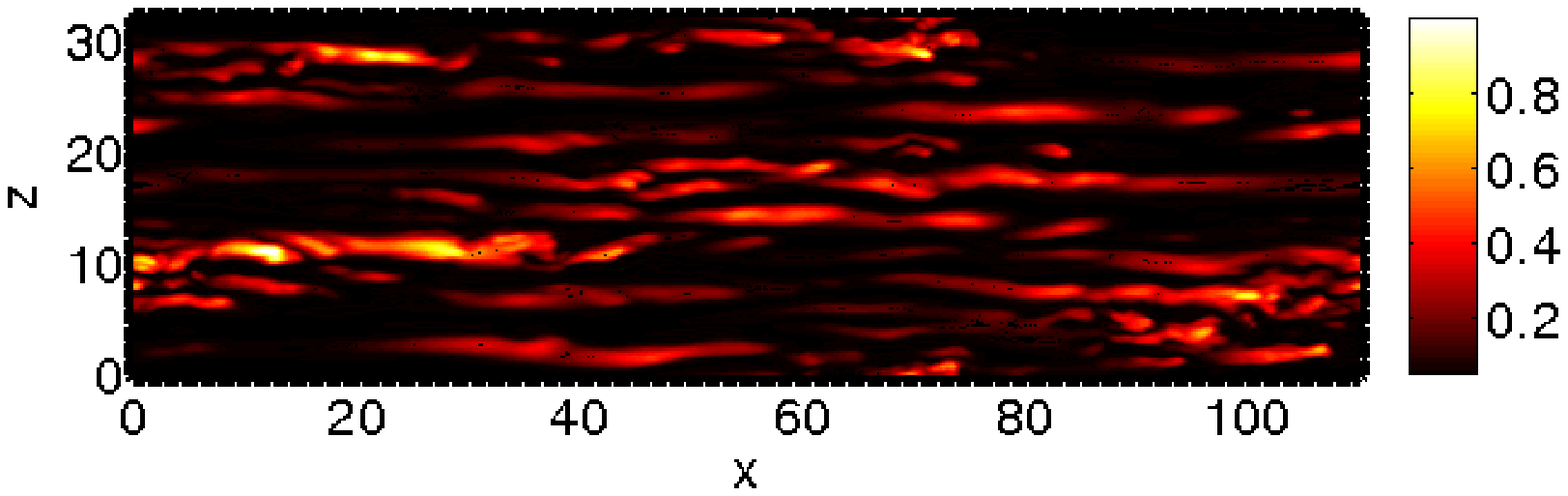}}
\rput(18,3.5){\includegraphics[width=4cm]{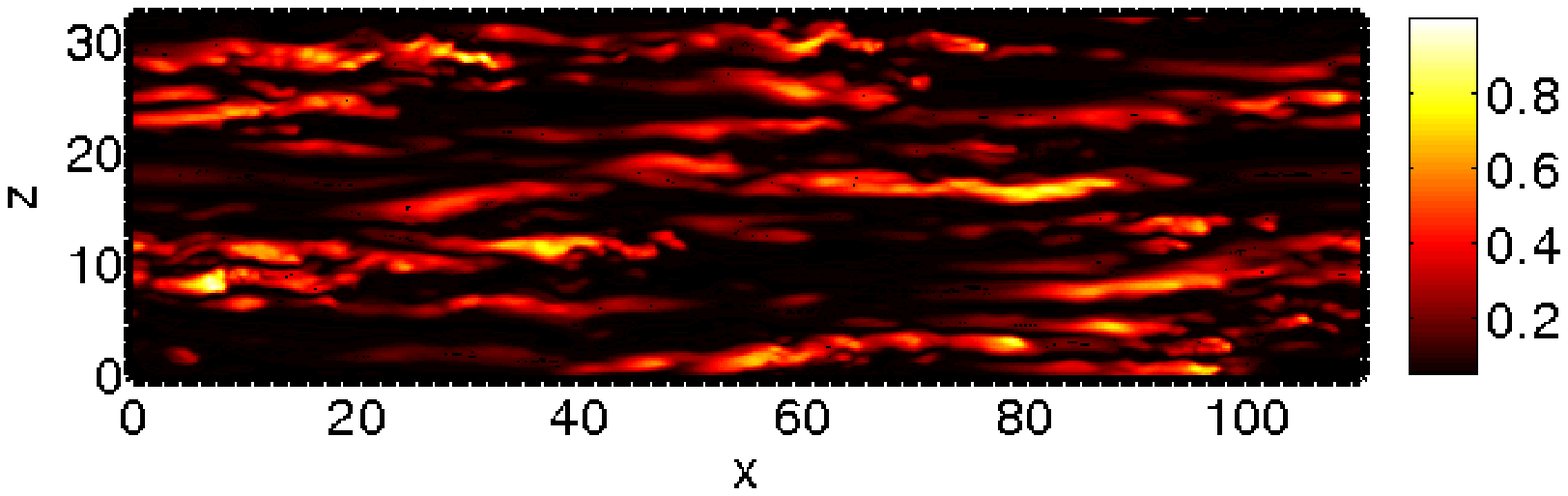}}
\rput(2,2){\includegraphics[width=4cm]{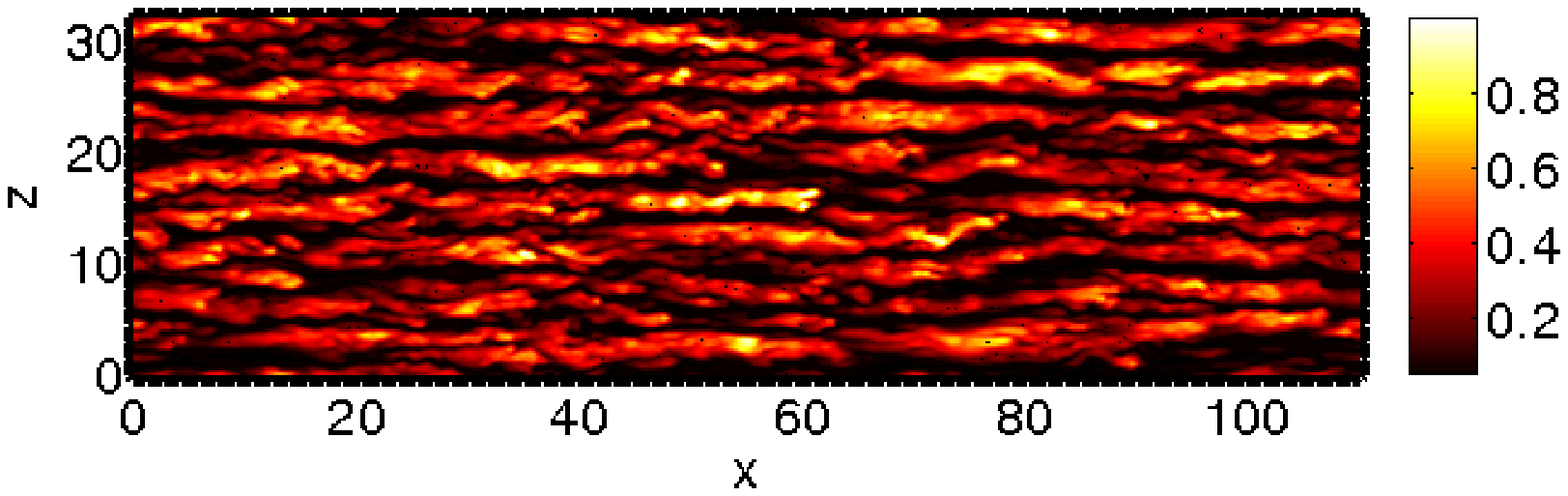}}\rput(6,2){\includegraphics[width=4cm]{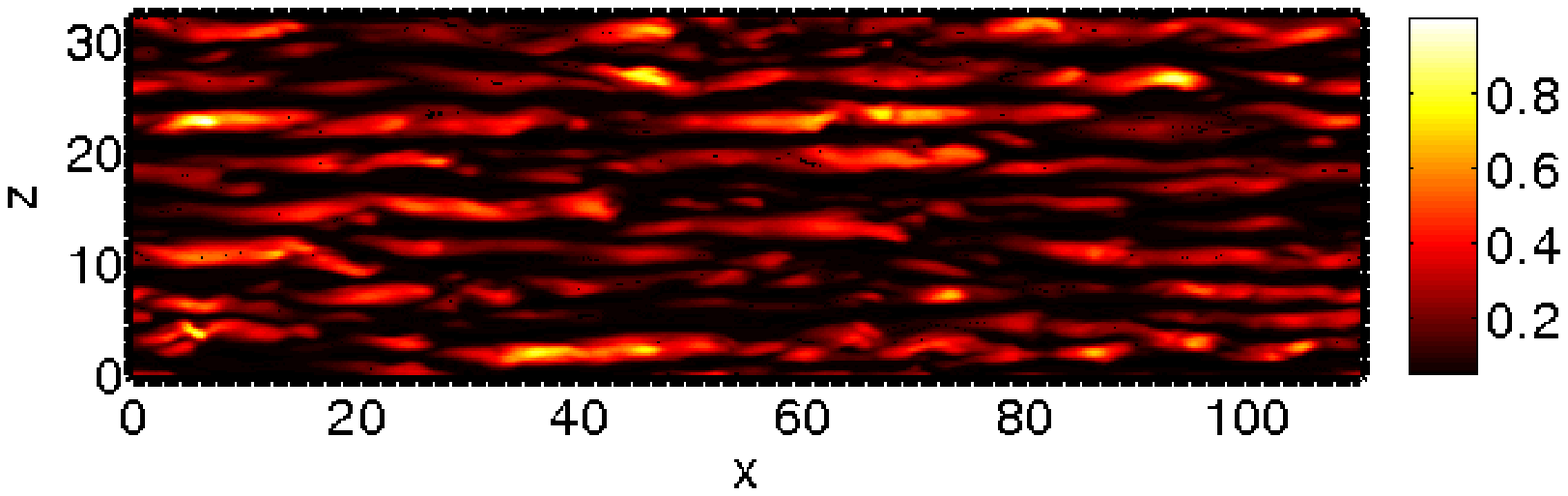}}
\rput(10,2){\includegraphics[width=4cm]{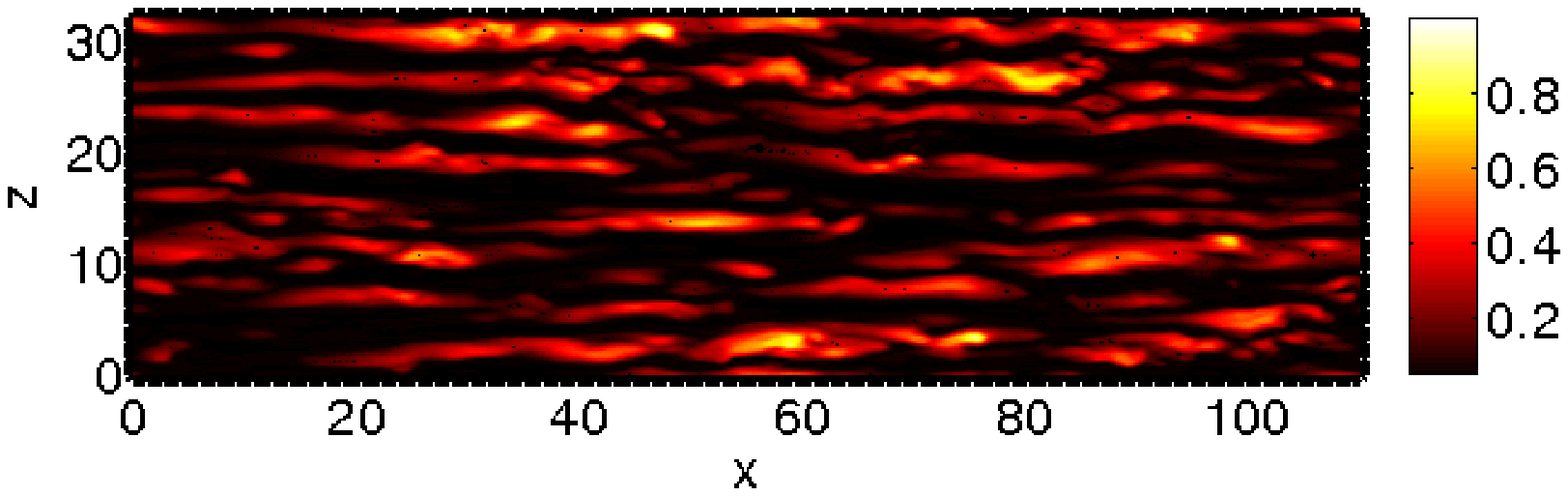}}\rput(14,2){\includegraphics[width=4cm]{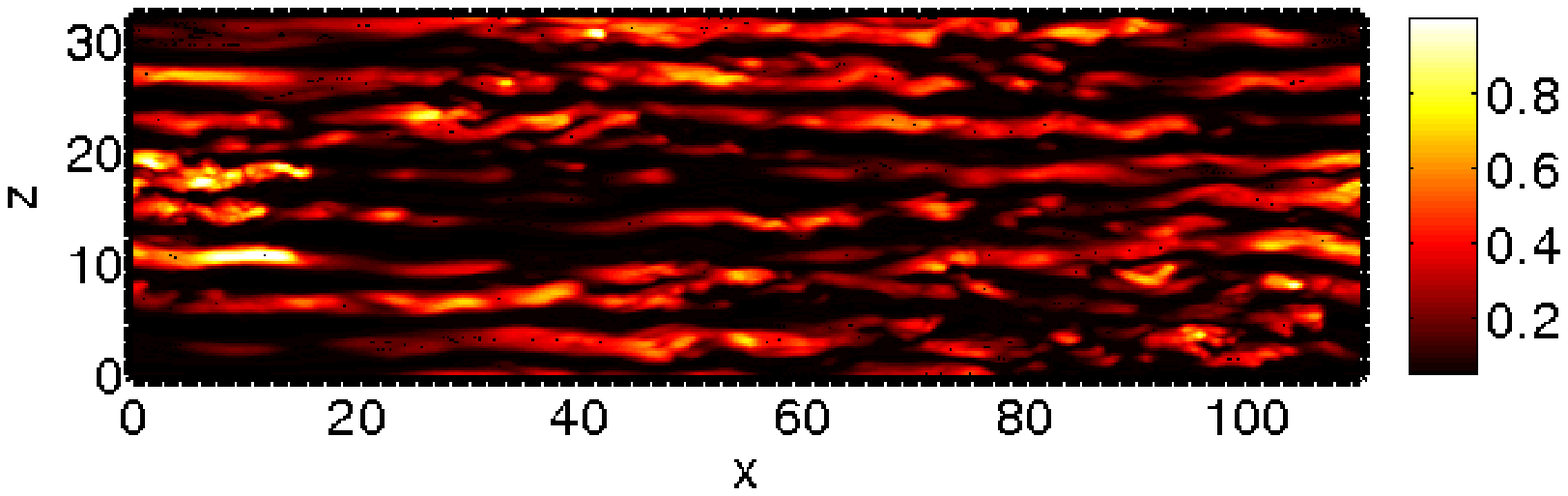}}
\rput(18,2){\includegraphics[width=4cm]{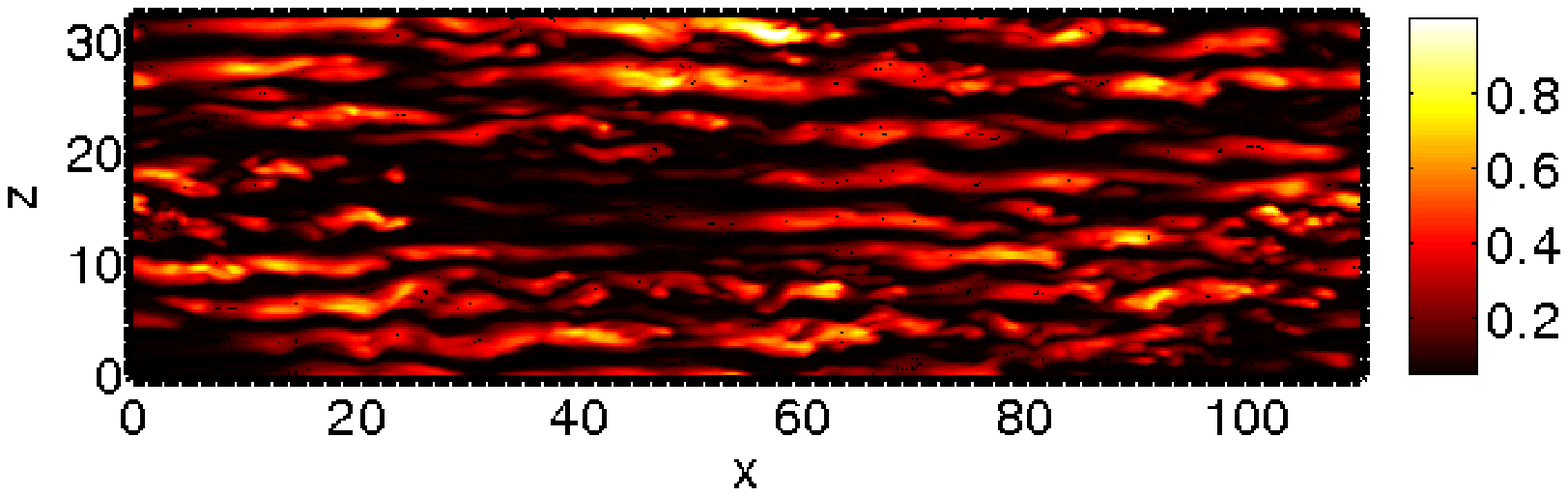}}
\rput(10,0.5){Snapshot of three trajectories}
\end{pspicture}
\end{flushleft}
\end{document}